\documentclass[fleqn,usenatbib]{mnras}

\usepackage{newtxtext,newtxmath}

\usepackage[T1]{fontenc}

\DeclareRobustCommand{\VAN}[3]{#2}
\let\VANthebibliography\thebibliography
\def\thebibliography{\DeclareRobustCommand{\VAN}[3]{##3}\VANthebibliography}

\usepackage{graphicx}	
\usepackage{amsmath}	
\usepackage{nccmath}
\usepackage{lipsum}
\usepackage{stfloats}
\usepackage{float}
\usepackage{lipsum}
\usepackage{enumitem}
\usepackage[T1]{fontenc}
\usepackage{adjustbox}
\usepackage{makecell}
\usepackage[french]{babel}
\usepackage{subfigure} 
\usepackage{color}

\title[Star-gas misalignments in EAGLE]{The origin of star-gas misalignments in simulated galaxies}

\author[Catalina I. Casanueva et al.]{
Catalina I. Casanueva,$^{1,2}$\thanks{E-mail: cicasanueva@uc.cl}
Claudia del P. Lagos,$^{3,4}$
Nelson D. Padilla$^{1,2}$, Thomas A. Davison$^{5,6}$
\\
$^{1}$Instituto de Astrofísica, Pontificia Universidad Católica de Chile, Av. Vicuña Mackenna 4860, Santiago, Chile\\
$^{2}$Centro de Astro-Ingeniería, Pontificia Universidad Católica de Chile, Av. Vicuña Mackenna 4860, Santiago, Chile\\
$^{3}$International Centre for Radio Astronomy Research (ICRAR), M468, University of Western Australia, 35 Stirling Hwy, Crawley, WA 6009, Australia\\
$^{4}$Australian Research Council Centre of Excellence for All-sky Astrophysics (CAASTRO), 44 Rosehill Street Redfern, NSW 2016, Australia\\
$^{5}$Jeremiah Horrocks Institute, University of Central Lancashire, Preston PR1 2HE, UK\\
$^{6}$European Southern Observatory, Karl-Schwarzschild-Strasse 2, D-87548 Garching bei Muenchen, Germany
}
\date{Accepted XXX. Received YYY; in original form ZZZ}

\pubyear{2021}

\begin{document}
\label{firstpage}
\pagerange{\pageref{firstpage}--\pageref{lastpage}}
\maketitle

\begin{abstract}
We study the origin of misalignments between the stellar and star-forming gas components of simulated galaxies in the EAGLE simulations. We focus on galaxies with stellar masses $\geq 10^9$ M$_\odot$ at 0$\leq$z$\leq$1. We compare the frequency of misalignments with observational results from the SAMI survey and find that overall, EAGLE can reproduce the incidence of misalignments in the field and clusters, as well as the dependence on stellar mass and optical colour within the uncertainties. We study the dependence on kinematic misalignments with internal galaxy properties and different processes related to galaxy mergers and sudden changes in stellar and star-forming gas mass. We find that galaxy mergers happen in similar frequency in mis- and aligned galaxies, with the main difference being misaligned galaxies showing a higher tidal field strength and fraction of ex-situ stars. We find that despite the environment being relevant in setting the conditions to misalign the star-forming gas, the properties internal to galaxies play a crucial role in determining whether the gas quickly aligns with the stellar component or not. Hence, galaxies that are more triaxial and more dispersion dominated display more misalignments because they are inefficient at realigning the star-forming gas towards the stellar angular momentum vector. 
\end{abstract}




\section{Introduction}\label{intro}

Understanding how galaxies formed and evolved to their observed state is one of the main interests of modern astrophysics. Studying the kinematical misalignments between the stellar and gaseous components can give us clues about the origin and the processes that affect gas accretion and possibly their galaxy assembly histories, thus playing a crucial role in our understanding of galaxy formation and evolution. 

Developments in spectrographs have led to the advent of Integral Field Units (IFUs) surveys, which provide spatially resolved spectra for galaxies in a way that allows the measurement of the angle between the rotational axes of stars and star-forming gas (SF gas) through the construction of two-dimensional kinematic maps of absorption and emission lines, respectively (e.g., \citealt{2011MNRAS.414..968D,2014MNRAS.443..485F}). 

Early IFU surveys focused on early-type galaxies (ETGs), including surveys such as SAURON \citep{2002MNRAS.329..513D} and ATLAS$^{\rm{3D}}$ \citep{2011MNRAS.413..813C}. \citet{2011MNRAS.414..968D} used 260 ETGs from ATLAS$^{\rm{3D}}$ to study the incidence of stellar-gas misalignments in ETGs and found that $\sim$ 36$\%$ of these galaxies have misaligned gas disks (i.e., the difference between rotational axes of stars and gas is greater than 30$^{\circ}$\footnote{This threshold angle comes from the typical error associated to this measurement being at most 30$^{\circ}$ both in observations (using the \textsc{fit$\_$kinematic$\_$pa} routine described in \citealt{2006MNRAS.366..787K}) and simulations \citep{2011MNRAS.414..968D}. Hence galaxies with differences in the stellar and gas kinematic position angle over 30$^{\circ}$ are clearly misaligned.}). This fraction increases for field ETGs, i.e., those outside the Virgo cluster or its vicinity. This high fraction of misalignments led to the hypothesis that the gas has an external origin, pointing to minor mergers as a possible main source of the misaligned gas in early-type galaxies (e.g., \citealt{2011MNRAS.414..968D,2012MNRAS.422.1835S}). 

Later, a new generation of IFUs such as SAMI \citep{2012MNRAS.421..872C,2015MNRAS.447.2857B}, MaNGA \citep{2015ApJ...798....7B}, and CALIFA \citep{2012A&A...538A...8S}, allowed the extension of these studies to late-type (LTGs) and lower mass galaxies. \citet{Chen_2016} used a representative sample of 1351 nearby galaxies from the MaNGA survey to find that $\sim$ 2$\%$ of blue star-forming galaxies have counter-rotating gas with a clear boost in star formation in their central regions. This suggests that these galaxies accrete abundant external gas that interacts with the pre-existing gas, driving gas into the central regions ($<$ 1 kpc) and leading to centrally concentrated rapid star formation. Further, \citet{Jin_2016} studied the properties of 66 galaxies with kinematically misaligned gas and stars from the MaNGA survey, finding that the misalignment fraction depends on physical galaxy properties such as stellar mass and specific star formation rate (sSFR). They concluded that the fraction of misaligned galaxies peaks at a stellar mass of $\sim10^{10.5}$ M$_{\odot}$ and increases with lower sSFR. They also found evidence that misaligned galaxies tend to reside in more isolated environments, in agreement with the findings of ATLAS$^{\rm{3D}}$ for ETGs. \citet{Duckworth_2018} used $\sim$ 900 central galaxies from the MaNGA survey to investigate the relationship of kinematically misaligned galaxies with their large-scale environment and with the halo assembly time. They found that large scale environment does not seem to contribute significantly to misalignments, while morphology appears to have a stronger effect since practically none of their misaligned galaxies are classified as LTGs by visual inspection. \citet{Bryant_2018} used $\sim$ 1200 galaxies from the SAMI galaxy survey and also demonstrated that morphology has the strongest correlation with the likelihood of star-gas decoupling rather than the local environment, and suggested that mergers are not the main driver for misalignments due to their low frequency. All this evidence supports the notion that the internal properties of galaxies are better predictors of misalignments than their environment.

Misalignments between the stellar and SF gas components are expected in the standard cosmological model, $ \Lambda $-Cold Dark Matter ($\Lambda$CDM). Within the $\Lambda$CDM framework, galaxies form by cooling and condensation of gas clouds within the potential wells of dark matter halos \citep{1978MNRAS.183..341W,1998MNRAS.295..319M}. At early times (i.e., during the linear stage of structure formation), baryons and dark matter are well mixed in such a way that both experience similar torques from the surrounding tidal field of protohalos (tidal torque theory TTT; e.g., \citealt{1969ApJ...155..393P,1970Afz.....6..581D,1984ApJ...286...38W}). In this paradigm, galaxies inherit the angular momentum content of the surrounding halo \citep{1980MNRAS.193..189F}. However, after the turnaround (i.e., after the decoupling of the protohalo region from the general expansion of
the universe), it is expected that the rotation of dark matter, gas, and stars may decouple due to cooling and other non-linear interactions. This decoupling has been evidenced in different numerical simulations (e.g., \citealt{2002ApJ...576...21V,2013MNRAS.429.3316B,2015A&A...584A..43P,2015ApJ...812...29T,10.1093/mnras/stw1286,10.1093/mnras/stw2945,2017ApJ...841...16D,10.1093/mnras/sty2513}).

Galaxy formation and evolution models can improve our interpretation of the observations by comparing the level of misalignment present in them. Simulation studies such as \citet{2012MNRAS.423.1544S}, who used 100 galaxies from the GIMIC cosmological gas-dynamical simulations \citep{2009MNRAS.399.1773C}, and \citet{2014MNRAS.443.2801P}, using the \textsc{sag} semi-analytic model \citep{2001MNRAS.328..726S,2006MNRAS.368.1540C,10.1111/j.1365-2966.2008.13456.x,2010MNRAS.408.2008T,2014MNRAS.443.2801P}, showed substantial evidence that the angular momentum of galaxies is not necessarily aligned with that of their host halos since accretion from either the cosmic web or mergers can be stochastic. \citet{2015MNRAS.448.1271L} used the semi-analytical model \textsc{galform} \citep{Cole_2002} coupled with a Monte-Carlo simulation \citep{2014MNRAS.443.2801P} to follow the angular momentum flips driven by matter accretion onto halos. Assuming that the only sources of misalignments are galaxy mergers, they found that only $\sim2-5\%$ of ETGs would show misaligned stars and gas, in tension with the observations of ATLAS$^{\rm{3D}}$. However, considering the smooth increase of gas as a driver of misalignment, they predicted that this fraction increased to $\sim46\%$. These results led to the conclusion that the high fraction of misaligned gas discs observed in ETGs is mostly due to smooth gas accretion (e.g., stochastic cooling from the hot halo of galaxies), which takes place after most of the stellar mass of the galaxy is in place and comes misaligned with respect to the stellar component. They also found that ETGs with high masses, low cold gas fractions and low star formation rates are more likely to display aligned cold gas and stellar components, in agreement with ATLAS$^{\rm{3D}}$ data. Furthermore, \citet{Starkenburg_2019} used the Illustris numerical simulations \citep{Nelson_2015} to study the origin of counter-rotation of gaseous disks in low mass galaxies (i.e., galaxies with a stellar mass between $\sim10^{9}-10^{10}$ M$_{\odot}$). They identified the importance of gas loss through black hole (BH) feedback and gas stripping during a fly-by passage across a more massive group environment in driving misalignments between stars and SF gas. More recently, \citet{2020ApJ...894..106K} using the Horizon-AGN simulation \citep{10.1093/mnras/stu1227} also found that ETGs are substantially more frequently misaligned than LTGs and that misalignment increases with decreasing gas fraction. They compared Horizon-AGN with SAMI, finding a significant discrepancy in the misalignment fraction for galaxies in dense environments. While no clear difference was found in the misalignment fraction between field and cluster environments of the SAMI observations, Horizon-AGN found a factor of three higher values in cluster galaxies regardless of morphology. \citet{2020arXiv201204659K} explored the origin of misalignments in Horizon-AGN, 
finding that 61$\%$ of the misalignments are not merger-driven, but they are related to interaction with nearby galaxies, interaction with dense environments, and secular evolution.
Finally, \citet{Duckworth2020a} using a combination of data from the MaNGA survey and MaNGA-like observations in IllustrisTNG100 \citep{Nelson2019TheIS}, also found a strong morphological dependence on the misalignment fraction and a crucial role of a significant gas loss in decoupling star-gas rotation. Furthermore, they found that central galaxies are more likely to exhibit misalignment than satellites and that misalignment at $z=0$ is correlated with the spin of the dark matter halo going back to $z=1$.  In a second paper \citep{Duckworth2020b} they found that misaligned low mass galaxies have experienced gas loss due to a higher energy injection through BH feedback, while
the origin of misalignment in massive quenched galaxies is more likely due to accretion of pristine gas or loss of enriched gas.

Despite this progress, the mechanisms for kinematic decoupling gas and stars are still not fully determined. We aim to identify which galaxies are more likely to display misalignments and what physical processes are responsible for these. We attempt to disentangle between external and internal factors that lead to misalignment or make galaxies more prone to misalignment.  In order to do this, we used the EAGLE cosmological hydrodynamical simulation \citep{2015MNRAS.446..521S,Crain_2015} to explore the dependence of kinematic misalignment on redshift, morphology and other galaxy properties. The unique combination of large cosmological volume and the sufficient resolution of EAGLE allows us to explore the origin of star-gas misalignments as well as to make reliable measurements of key galaxy quantities, such as angular momentum, morphology, and star formation activity. EAGLE is well suited for this study as it broadly reproduces the morphological diversity of galaxies \citep{2015MNRAS.446..521S,2017MNRAS.464.3850L,2019MNRAS.483..744T}, the fraction of star-forming and passive galaxies \citep{2015MNRAS.450.4486F,2019MNRAS.487.3740W}, and the stellar kinematic properties of galaxies and their dependence with stellar mass and environment \citep{2018MNRAS.476.4327L,2020MNRAS.494.5652W}.

This paper is organised as follows. In Section \ref{sec2:EAGLE} we describe the EAGLE simulation and our methods to study the origin of stars-gas misalignments in the simulation. In Section \ref{qual} we qualitatively compare misaligned galaxies in EAGLE with observed data from the SAMI survey. In Sections \ref{charac} and \ref{proc}, we examine and discuss possible internal properties and physical processes related to misalignments, including two study cases. Finally, in Section \ref{conclu} we summarise our results.

\section{The EAGLE simulations}\label{sec2:EAGLE}

EAGLE (Evolution and Assembly of GaLaxies and their Environments project; \citealt{2015MNRAS.446..521S,Crain_2015}) is a suite of hydrodynamical simulations that follow the formation, assembly, and evolution of galaxies from $z=127$ until $z=0$. These simulations were constructed assuming a $\Lambda$CDM cosmology with $\Omega_{\Lambda}=0.693$, $\Omega_{m} = 0.307$, $\Omega_{b}=0.04825$, $\sigma_{\rm{8}}=0.8288$ and $h = 0.6777$, consistent with Planck measurements \citep{refId0}.

The suite was simulated with a modified version of \textsc{Gadget}-3 N-Body Tree-PM smoothed particle hydrodynamics (SPH) code, which is an updated version of \textsc{Gadget}-2 (described in \citealt{2005MNRAS.364.1105S}). This version includes modifications to the hydrodynamics algorithm and the time-stepping criteria \citep{10.1093/mnras/stv2169} and incorporates sub-grid modules that govern the phenomenological implementation of physical processes that act on scales below the resolution limit of the simulations.

The sub-grid physics used in EAGLE include star formation \citep{2008MNRAS.383.1210S}, heating and cooling of gas \citep{2009MNRAS.393...99W}, stellar mass losses due to stellar evolution \citep{Wiersma_2009b}, energy feedback from massive stars \citep{2012MNRAS.426..140D}, gas accretion onto and mergers of supermassive black holes \citep{2009MNRAS.398...53B,2015MNRAS.454.1038R}, active galactic nuclei (AGN) feedback \citep{2015MNRAS.454.1038R} and photoionisation due to an evolving X-rays-to-UV background following \citep{2001cghr.confE..64H}. The model parameters were calibrated to reproduce the observed galaxy stellar mass function at $z \sim$ 0 and the black hole mass-stellar mass relation at $z \sim$ 0. In addition, the dependence of the stellar feedback energy on the gas density was introduced to reproduce the relation between galaxy mass and size of star-forming galaxies at $z \sim$ 0.1. \citet{Crain_2015} provide a comprehensive overview of the calibration process.

EAGLE galaxies and their host halos are identified by a multi-stage process, starting with the implementation of the friends-of-friends (\textsc{fof}) algorithm \citep{1985ApJ...292..371D} to the dark matter particle distribution. This algorithm links the particles together if their distance lies below the linking length of $b\sim0.2$ times the mean inter-particle separation. Gas, stars and black holes particles are associated with the \textsc{fof} group of their nearest neighbour dark matter particle (if there is any). After this, the \textsc{subfind} algorithm \citep{2009MNRAS.399..497D} is used to identify locally overdense self-bound substructures or subhalos within the full particle distribution of \textsc{fof} halos. Galaxies are then associated with subhalos, and hence in this work as in all the publications done with EAGLE, we assume galaxies and subhalos are equivalent.

Galaxies form and evolve within their host halos, so tracing them across snapshots is analogous to tracing the evolution of their host halos via merger trees \citep{2017MNRAS.464.1659Q}. The EAGLE merger trees were constructing by applying the \textsc{D-Trees} algorithm \citep{Jiang_2014} to the \textsc{subfind} subhalo catalogues across all simulation snapshots (i.e., across all simulation outputs at different redshifts). In essence, the algorithm links a subhalo with its descendant across more than two consecutive snapshots. It identifies a subhalo descendant by tracing where the majority of the most bound particles are located at the next output time. 

In this study, we use the EAGLE largest reference simulation, labelled as ‘Ref-L100N1504’ in \citet{2015MNRAS.446..521S}. This model was run in a cosmological volume of 100 comoving Mpc on a side and achieves a resolution of
0.7 proper kpc (for the gravity). The initial conditions of the simulation are generated using the \textsc{panphasia} multiresolution phases of \citet{2013arXiv1306.5771J}, such that the initial gas particle mass is $m_{\rm{GAS}}=1.81\times10^{6}$ M$_{\odot}$ and the mass of dark matter particles is $m_{\rm{DM}}=9.70 \times 10^{6}$ M$_{\odot}$. There is initially an equal number of $1504^{3}$ baryonic and dark matter particles. The properties of the model particles were recorded for 29 snapshots between redshifts 20 and 0, which translates into time span ranges between snapshots from $\approx$ 0.3 to 1 Gyr.
To avoid biasing our results by the coarse time cadence of the snapshots, we also make use of the EAGLE ``snipshots'', which are lean outputs of the simulation with a time cadence of $\sim 30-80$~Myr. Merger trees were also produced for these outputs (see \citealt{2017MNRAS.464.4204C}) and the gas particle data has enough information to compute misalignment fractions, and several of the intrinsic properties of galaxies we introduce in $\S$~\ref{sec:prop}.

\begin{figure*}
 \centering
 \includegraphics[width=0.9\textwidth]{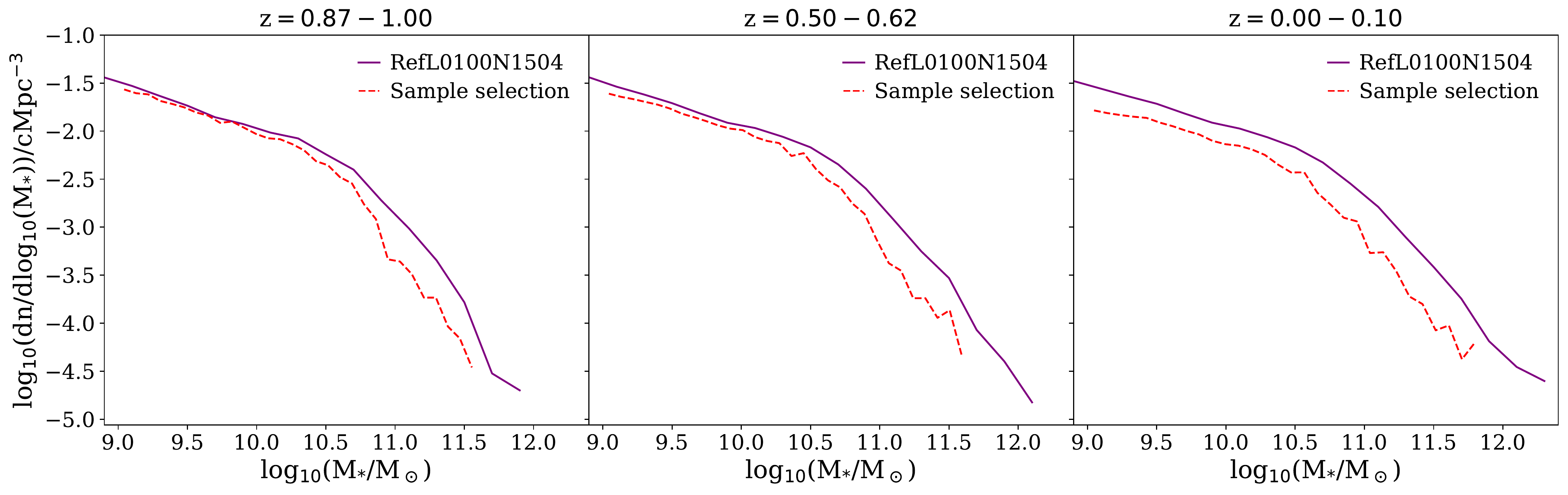}
 \caption{Stellar mass function of all galaxies in the RefL0100N1504 simulation (solid purple lines) in comparison with our sample selection (dashed red lines), at the three redshift ranges as labelled in each panel.}\label{SMF}
\end{figure*}

\subsection{Sample selection}\label{sampleselection}

As mentioned in the previous section, we used galaxies from the Ref-L100N1504 simulation (hereafter referred to as the EAGLE simulation). In order to avoid resolution problems and make our results comparable to observations, we work with a sample of galaxies under the following limits:
\renewcommand\labelitemi{{\boldmath$\cdot$}}
\begin{itemize}
        \item Stellar mass $\geqslant$ $\mathrm{10^9}$ M${_\odot}$ (corresponding to $\approx 1000$ stellar particles),
        \item Number of star-forming gas particles $\geqslant$ 20,
        \item The centres of mass of the stars and the star-forming gas should be at a distance < 2 kpcs. 
\end{itemize}

Star-forming gas particles are all those that have a star formation rate (SFR) > 0, and therefore are the closest to H-alpha emitting gas, which is what is broadly used to measure misalignments with the stellar component (e.g., \citealt{Bryant_2018}).

The number of SF gas particles was chosen such that the sample included not only galaxies on the main sequence of star formation in the SFR-stellar mass plane but also below it, well into the passive population. This is important as observations suggest elliptical galaxies, which tend to be more passive than late-type galaxies, are more likely to display misalignments \citep{Duckworth_2018,Bryant_2018}. Nevertheless, we tested increasing the lower limit to 100 SF gas particles, and our results remain unchanged.

The limit between the distance of the centres of mass of stars and gas was considered because, in observations, misalignments are usually studied when gas is well associated with a galaxy. \citet{2019MNRAS.485.5715T} presented an analysis of the resolved star formation main sequence and metallicity-mass relation. In the example synthetic cubes they show in Appendix D, it is clear that some galaxies in EAGLE have clouds of star-forming gas offset with respect to stars. Our criteria avoid the inclusion of those galaxies.

A possible concern is whether the angular momentum vector of stars and gas can be well measured with the number of particles above. This can become problematic for very dispersion dominated galaxies in particular. We use the stellar spin parameter, $\lambda_R$ \citep{2007MNRAS.379..401E}, to quantify how many of our galaxies would fall in the dispersion dominated category (see \citealt{10.1093/mnras/sty489} for details on how $\lambda_R$ is computed in EAGLE galaxies). We select dispersion dominated galaxies as those having an edge-on $\lambda_R<0.1$ and find that only 0.03$\pm$0.03$\%$ of our sample falls in this category. If we only look at misaligned galaxies, the fraction is 0.23$\pm$0.23$\%$. This shows that most of them have a relatively well-defined kinematic major axis.

Fig. \ref{SMF} shows that the shape of the stellar mass function of our sample is similar to that of the total galaxy population. This means that our selection is not biased towards specific galaxy masses.

\subsection{Internal galaxy properties}\label{sec:prop}

In what follows, we define the main properties analysed in this study. We include morphological proxies and measurements related to the star-formation activity to investigate if these could determine whether the star-forming gas and the stellar component of the galaxy become misaligned or remain aligned for long enough to detect it in the snapshots of the simulation.

\begin{itemize}

    \item \textbf{Star-forming gas}: Gas particles are eligible to form star particles if they have cooled to reach densities greater than the metallicity-dependent threshold described in \citet{2015MNRAS.446..521S} (equation 2):
    
\begin{ceqn}
\begin{align}
 n^{*}_{\rm{H}} = 10^{-1} \rm{cm}^{-3} \left(\frac{Z}{0.002}\right)^{-0.64},
\end{align}
\end{ceqn}   

\end{itemize}
where Z is the gas metallicity. 

\begin{itemize}

    \item \textbf{Star-forming gas metallicity} ($Z_{\rm{SF}}$) and \textbf{stars metallicity} ($Z_{*}$): Mass fraction in metals of cold gas and stars, used to describe the abundance of elements present in these galaxy components that are heavier than helium. A ratio between these properties can be used to indicate galaxy mergers or recent gas accretion \citep{2020MNRAS.495.2827C}.
    If galaxies have acquired their gas content via minor mergers, one would expect the gas-to-stellar metallicity ratio to be low compared to galaxies that have not had mergers.
   \newline
    \item \textbf{Specific star formation rate} (sSFR): The star formation rate per unit stellar mass. Star formation rate is defined as the mass of stars formed per unit time before any mass loss due to winds and supernovae. This is therefore used as an indication of the typical efficiency of stellar mass growth in galaxies and serves as a useful proxy for stellar mass growth timescale \citep{2014MNRAS.444.2637M}. We employ the total stellar mass and the total SFR of the galaxy, i.e., the measurements within a 3D aperture of 70 pkpc.
   \newline
    \item \textbf{Gas fraction} ($f_{\rm{gas}}$): Ratio between the total star-forming gas mass and the total stellar mass.
   \newline
\item \textbf{Stellar co-rotating kinetic energy fraction} ($\kappa_{\rm{co}}$): Fraction of stellar particle's total kinetic energy ($K$) involved in ordered co-rotation ($K^{\rm{rot}}_{\rm{co}}$):

\begin{ceqn}
\begin{align}
 \kappa_{\rm{co}}=\frac{K^{\rm{rot}}_{\rm{co}}}{K} = \frac{1}{K}\sum_{i,\textit{L}_{\rm{z,i}}>0}\frac{1}{2}m_{\rm{i}}\left(\frac{L_{\rm{z,i}}}{m_{\rm{i}}R_{\rm{i}}}\right)^{2},
\end{align}
\end{ceqn}
\end{itemize}
where the sum is over all co-rotating stellar particles within a spherical radius of 30 pkpc centered on the minimum of the potential, $m_{i}$ is the mass of each stellar particle, $K = \sum_{\rm{i}}\frac{1}{2}m_{\rm{i}}v_{\rm{i}}^{2}$ the total kinetic energy in the centre of mass frame, $L_{\rm{z,i}}$ the particle angular momentum along the direction of the the total angular momentum of the stellar component of the galaxy and $R_{\rm{i}}$ is the 2-dimensional radius in the plane normal to the rotation axis \citep{2017MNRAS.472L..45C}.

\citet{2017MNRAS.472L..45C} found that dividing the EAGLE population by applying a threshold in $\kappa_{\rm{co}}$ of $\approx$ 0.4 is a good way of separating between the `blue cloud' of disky star-forming galaxies ($\kappa_{\rm{co}} > 0.4$) and the `red sequence' ($\kappa_{\rm{co}} < 0.4$) of spheroidal passive galaxies in the galaxy colour-stellar mass diagram. 

\begin{itemize}
\item \textbf{Disc-to-total stellar mass ratio} ($D/T$): The stellar mass fraction that is supported by rotation. This can be considered as the ‘disc’ mass fraction and is calculated assuming that the bulge component has no net angular momentum (e.g., \citealt{Crain_2010,Clauwens_2018}). Thus, the bulge mass can be estimated as twice the mass of stars that are counter-rotating with respect to the galaxy, and the disc-to-total mass fraction ($D/T$) is the result of subtracting the bulge-to-total ($B/T$) mass fraction, i.e.,:
\begin{ceqn}
\begin{align}
 \frac{D}{T} = 1-\frac{B}{T} = 1-2\frac{1}{M_{*}}\sum_{i,L_{\rm{z,i}}<0}m_{\rm{i}},
\end{align}
\end{ceqn}
 
 \end{itemize}
 where the sum is over all counter-rotating $(L_{\rm{z,i}}<0)$ stellar particles within 30 pkpc.

\citet{Thob_2019} inferred that dividing the population with a threshold of
$D/T \sim0.45$ provides a means of separating the star-forming from passive galaxy populations. Note that by definition, we do expect $D/T$ to be well correlated with $\kappa_{\rm{co}}$.
\begin{itemize}
\item \textbf{Triaxiality} ($T$): Stellar triaxiality from the iteratively reduced inertia tensor. This parameter characterises an ellipsoid that models the spatial distribution of stars in galaxies:
\begin{ceqn}
\begin{align}
 T = \frac{a^2-b^2}{a^2-c^2},
\end{align}
\end{ceqn}
\end{itemize}
where $a$, $b$ and $c$ are the moduli of the major, intermediate and minor axes, respectively

Low and high values of $T$ correspond to oblate and prolate ellipsoids, respectively. Oblate systems exhibit axisymmetry about the minor axis, while prolate galaxies are characterised by an intermediate axis that
is significantly shorter than their major axis, and thus resemble cigars. Galaxies are classified as oblate when $T < 1/3$. 
\begin{itemize}
\item \textbf{Stellar velocity anisotropy} ($\delta$): Describes the anisotropy of the galaxy's velocity dispersion:
\begin{ceqn}
\begin{align}
\delta = 1-\left(\frac{\sigma_{\rm{z}}}{\sigma_{0}}\right)^{2},
\end{align}
\end{ceqn}
\end{itemize}
where $\sigma_{\rm{z}}$ is the velocity dispersion in the plane of the rotation axis (where the latter is the unit vector parallel to the total angular momentum vector of all stellar particles within 30 pkpc), and $\sigma_{0}$ is the velocity dispersion in the ‘disc plane’, i.e., the plane normal to the z-axis, selected in order to recover an estimate of the line-of-sight velocity dispersion to make dispersion measurements comparable with observations.

Values of $\delta > 0$ indicate that the velocity dispersion is primarily contributed by disordered motion in the disc plane, i.e., that is defined by the intermediate and major axes, rather than disordered motion in the direction of the minor axis \citep{Thob_2019}.
\begin{itemize}
\item \textbf{Stellar velocity rotation-to-dispersion ratio} ($v_{\rm{rot}}/\sigma_{0}$): This ratio is often used as a kinematical diagnostic because both, the rotation velocity and the velocity dispersion, can be estimated from spectroscopic observations of galaxies \citep{10.1093/mnras/stx1751}. 

\citet{Thob_2019} infer that division about a threshold of $v_{\rm{rot}}/\sigma_{0} \sim0.7$ separates the disky star-forming and spheroidal passive galaxy populations with a similar efficacy to the $\kappa_{\rm{co}} = 0.4$ threshold defined  by \citet{2017MNRAS.472L..45C}.

\end{itemize}

\begin{figure*}

\hspace*{0cm}\includegraphics[width=15.8cm]{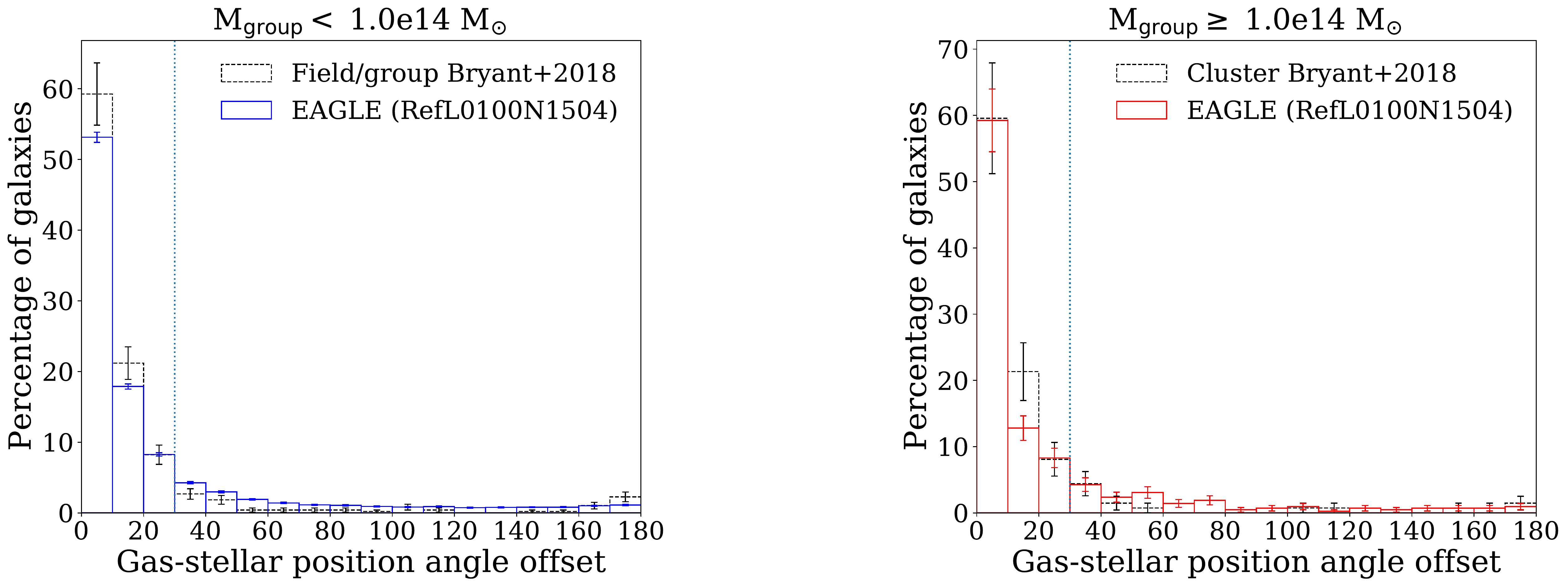}
\caption{Distribution of the angles between the rotation axis of the stars and star-forming gas for galaxies that reside in groups of mass $<10^{14}$ M$_{\odot}$ (left) and galaxies that reside in groups of mass $>10^{14}$ M$_{\odot}$ (right). Blue dotted lines mark a PA offset $= 30^{\circ}$ to separate between aligned and misaligned galaxies. Dotted and solid histograms show the distribution for galaxies in SAMI \citep{Bryant_2018} and EAGLE, respectively. Error bars in the EAGLE simulations show Poisson errors.}\label{histangle2}

\end{figure*}

Although we do expect most of these quantities to be closely correlated, there are some subtle differences such as intrinsic scatter, non-linearities, and multivariate distributions. Hence, it is important to understand if some are better correlated with the probability of finding misalignments.

In this work, we use 2D angles between the rotational axes of SF gas and the stars, measured for particles within a radius that encompasses half of the total stellar mass. This is motivated by the fact that observations typically measure these angles within an effective radius (encompassing half of the optical light).

Observational data are projected to the plane perpendicular to the line of sight. Thus, in order to mimic this projection effect, in this work we will use the projected misalignments from the position angles (PA offset or misalignment angle) rather than the three-dimensional misalignments. The kinematic PA is defined as the position angle of the axis passing through the stellar centre of mass along which the line-of-sight rotating velocity reaches a maximum and is measured counterclockwise with respect to a specific axis $x$ (the choice of $x$ is arbitrary because we are interested in the kinematic misalignment which involves the difference of two angles).

\subsection{Internal and external physical processes}\label{sec:ext}

We track the evolution of galaxies in EAGLE through time via the galaxy merger trees. Using the parameters \textbf{GalaxyID} and \textbf{DescendantID} of the EAGLE database tables, we follow the main progenitor branch, which for any snapshot is defined as the branch with the largest total mass summed across all the earlier snapshots. This allows us to identify when a misalignment takes place in galaxies and how other galaxy properties are changing. 

By comparing the galaxy status between two consecutive snapshots, we will explore the following physical processes that could lead to kinematic misalignment:
\begin{itemize}
\item \textbf{Galaxy mergers:} We will consider that there was a merger of two or more galaxies if they have the same DescendantID and if the merger ratio is greater than 0.1. This limit allows us to ensure that we do not include dark-matter-only subhalo mergers (which are the most abundant) or very minor mergers which are generally not well resolved. The merger ratio is defined as the ratio between the stellar masses of the two most massive galaxies involved in the merger, and it is computed in a way that it is strictly $\leq$ 1.

\item \textbf{Abrupt increase/decrease in stellar mass:} We will study sudden increases/decreases in stellar mass, i.e., abrupt changes between two consecutive snapshots that are not product of a merger.  We will divide these increases/decreases into two groups: between 5$\%$ and 10$\%$ (labelled as `$\uparrow$ / $\downarrow$ $5\%$') and by $\geq 10\%$ (labelled as `$\uparrow$ / $\downarrow$ $10\%$').

\item \textbf{Abrupt increase/decrease in star-forming gas mass:} Same as the previous cause but for star-forming gas.

\item Combinations between the last two points, i.e., when sudden changes in stellar mass and SF gas mass co-occur. 
\end{itemize}

The processes described above can be divided into internal and external processes. Galaxy mergers are clearly external, while the abrupt increases or decreases of star-forming gas or stars could be internal if they are due to feedback or star-formation, or external if they are due to a perturbation (for example, by ram pressure stripping or by a close fly-by).

\section{Physical drivers of misalignments in simulated galaxies}\label{metodo}

This section analyses whether there are systematic differences in the population of aligned vs misaligned galaxies in terms of their internal and external properties.

First, we compare our results with observations. Then, we characterise the aligned and misaligned populations by the internal galaxy properties described in Section \ref{sec:prop}. Finally, we study how the different physical processes described in Section \ref{sec:ext} are related to misalignments.

\subsection{Comparison with observations}\label{qual}

\begin{figure}
\begin{center}

\includegraphics[width=8.5cm]{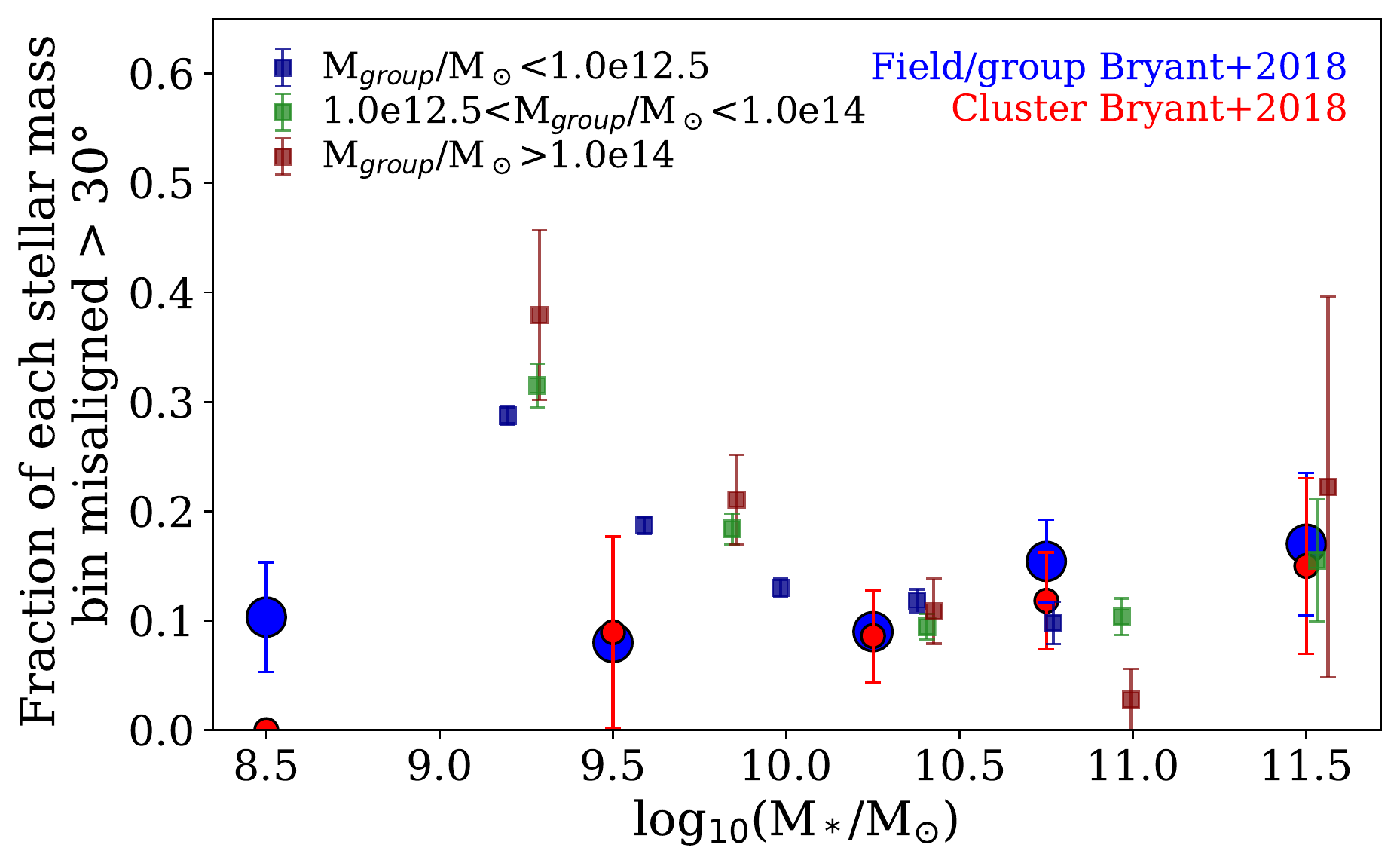}\\
\vspace{0.6cm}
\includegraphics[width=8.5cm]{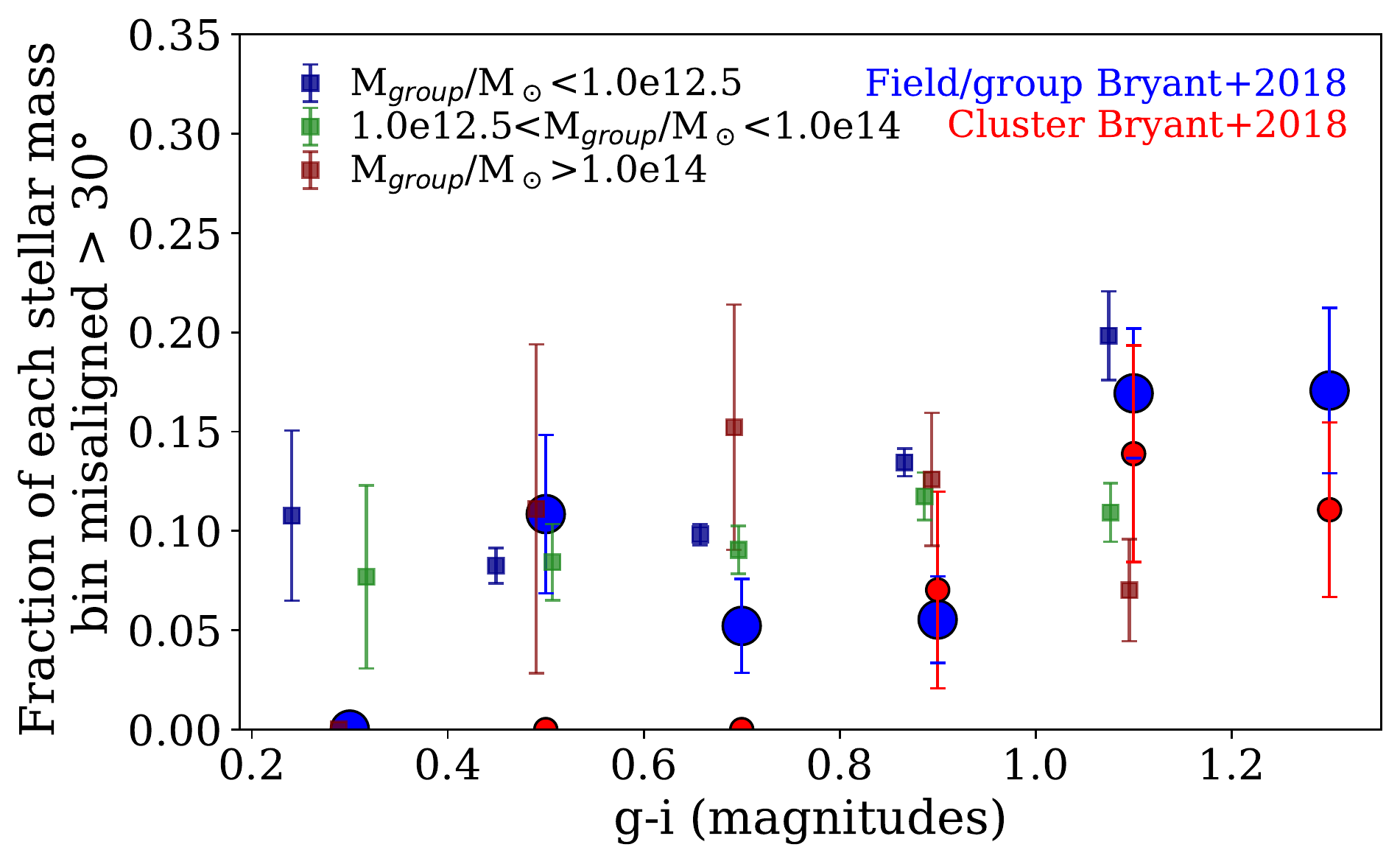}
\caption{Fraction of galaxies that are misaligned (PA offset $>30^{\circ}$) in bins of stellar mass (top panel) and in bins of g-i colour (bottom panel). The circles are the results obtained with EAGLE simulated galaxies and the squares the results of \citet{Bryant_2018}, using SAMI. Our sample is separated into three ranges in halo mass, as labeled. Simulated data errors correspond to Poisson uncertainties.}\label{smygi}
\end{center}
\end{figure}

In order to make a comparison with integral field spectroscopic observations from the SAMI survey, we used EAGLE simulated galaxies at $z=0.00$ and $z=0.10$ that satisfy the sample selection described in Section \ref{sampleselection}. It is important to remind the reader that, although PA is measured differently in simulations and observations, our results capture the fact that this quantity is usually calculated within an effective radius and in projection.

Fig. \ref{histangle2} shows the distribution of the PA offsets for two ranges of group mass in which galaxies reside. We find that the fraction of misaligned galaxies in both samples is almost
the same ($\sim20\%$), suggesting that the group mass does not influence the chance of being misaligned. \citet{Bryant_2018}, using a sample of 1213 galaxies at $z<0.1$ from the SAMI Galaxy Survey, also found the same, although with a lower misalignment fraction ($\sim11\%$). Furthermore, they found a small peak at PA offset $\sim$ 180° (of $\sim$ 4$\%$), which is not clearly visible in EAGLE, nor was it reproduced by Horizon-AGN according to \citet{2020ApJ...894..106K}. It is possible that the resolution of the simulations causes this small peak to dissolve faster than in real galaxies. Future simulations of similarly large boxes but at higher resolution are required to confirm this hypothesis.

Fig.~\ref{smygi} shows the fraction of galaxies with PA offset > $30^{\circ}$ in bins of stellar mass and g-i colour. We find a decrease in the misaligned fraction with increasing stellar mass for galaxies with stellar mass less than $10^{10.8}$ M$_{\odot}$. This trend is not observed in \citet{Bryant_2018}, but in general, the fractions of misaligned galaxies predicted by EAGLE are similar to those obtained with observational data for M$_{*}>10^{10}$M$_{\odot}$. We also show the effect of the environment in EAGLE by showing the fraction of misaligned galaxies in three bins of halo mass. We find that a clear environmental effect is only seen for stellar masses less than $10^{10}$ M$_{\odot}$, where groups and clusters display a larger fraction of misaligned galaxies. These trends are not clear in observations, but this could be partially due to the fact that observations do not distinguish between field and groups and instead place all galaxies in the field and groups with masses $<10^{14}$ M$_{\odot}$ in the same sub-sample. Also, we need to consider that the stellar mass range in SAMI is different from our simulated sample.
The fact that a trend with environment and the presence of misalignments is only seen in galaxies with stellar masses below $10^{10}\,\rm M_{\odot}$ is not necessarily surprising, as both observations (e.g. \citealt{Peng2010}) and simulations (e.g. \citealt{Trayford2016,2018MNRAS.476.4327L,Cochrane2018,2019MNRAS.487.3740W}) have shown that environment is relevant in determining galaxy morphology and their star formation activity (all properties that we find correlate with the presence of misalignments in $\S$~\ref{charac}) in low mass galaxies, while massive galaxies have these properties primarily correlating with stellar mass rather than environment.

We find a relatively flat misalignment fraction as a function of galaxy colour in agreement with \citet{Bryant_2018}, within the uncertainties. We see that the level of misalignments in galaxies of fixed colour are similar between clusters, groups and the field. A very weak trend is seen between the galaxy colour and the presence of misalignments that is most apparent at low halo masses, $>10^{12.5}\,\rm M_{\odot}$, with redder galaxies having slightly higher misalignment fractions than blue galaxies. If we instead look at galaxy colour at fixed stellar mass, we find a clearer correlation with red galaxies having more misalignments (not shown here). As we will see in the next section, there is a correlation with the specific star formation rate at fixed stellar mass (Fig.~\ref{pass-act}), which goes in the same direction as the one shown here for colour, but is much stronger.

\begin{figure}
\begin{center}

\includegraphics[width=8cm]{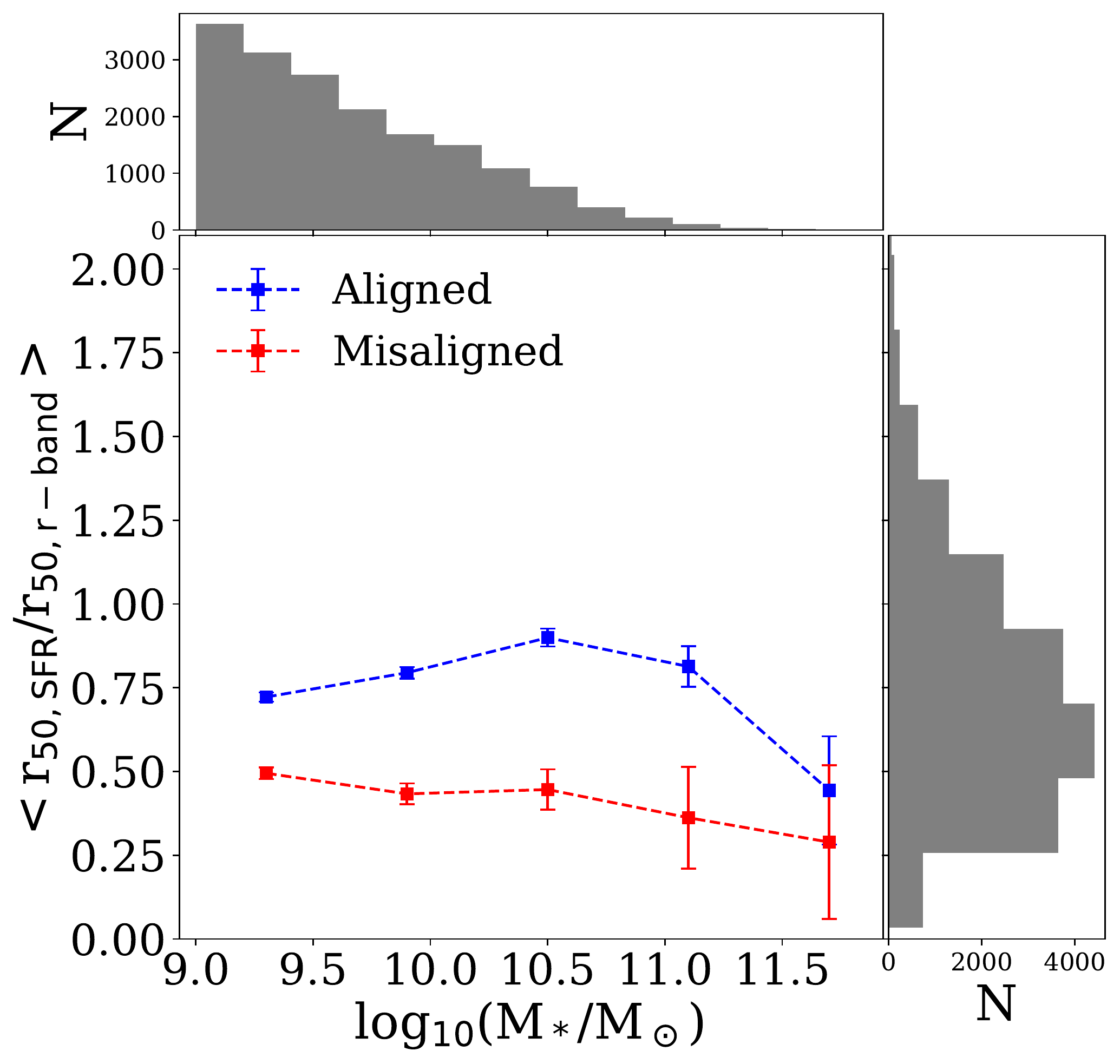}\\
\vspace{0.6cm}
\includegraphics[width=8cm]{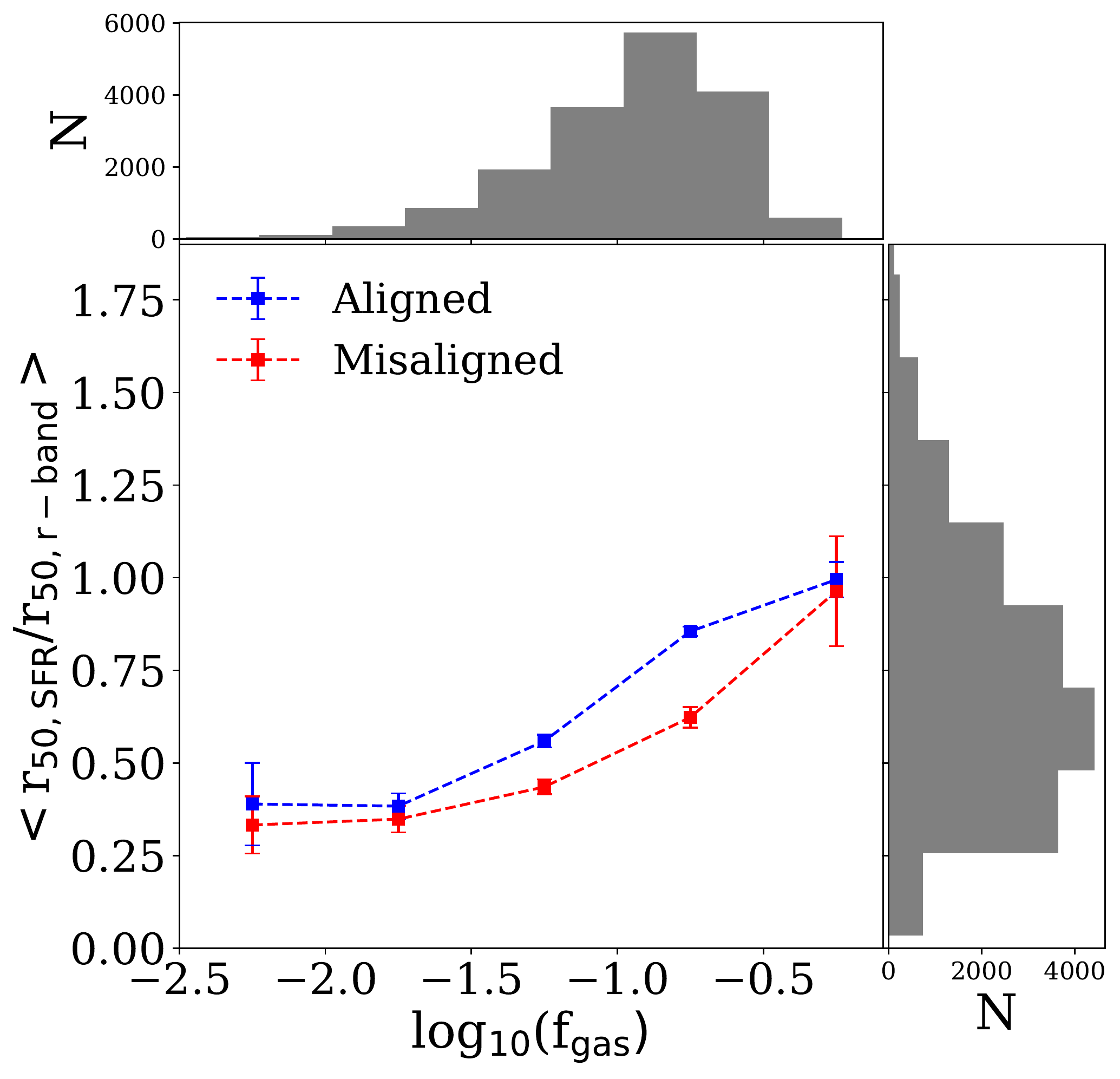}
\caption{Median ratio between the half-star-forming gas mass and half-stellar mass radii for each bin of stellar mass and gas fraction (top and bottom panels, respectively), with histograms showing the distribution of the data in each axis, for galaxies at $z=0.0-0.1$. In red we see the distribution for misaligned galaxies and in blue for aligned galaxies. Errors correspond to Poisson uncertainties.}\label{r50m}
\end{center}
\end{figure}

We explored additional galaxy properties in search of possible correlations with misalignment fractions that could be easily tested. Fig. \ref{r50m} shows the ratio between the half-star-forming gas mass and half-stellar mass radii for galaxies that are misaligned and those that are aligned as a function of stellar mass (upper panel) and as a function of the gas fraction (bottom panel). These measurements of radii are performed in the 3D distribution of stellar and star-forming gas particles. We find that at fixed stellar mass, misaligned galaxies have star-forming gas components that are significantly more compact with respect to the stellar component than aligned galaxies (see the upper panel of Fig. \ref{r50m}). 

Most of the trends seen in the SF gas and stellar mass radii ratio are driven by the differences in the gas fraction of the misaligned vs aligned galaxies, as gas-poor galaxies on average tend to have smaller half-mass radii. However, we find that this is not the whole story, as at fixed gas fraction misaligned galaxies still display a trend of smaller half-mass radii ratio than aligned galaxies (see bottom panel of Fig. \ref{r50m}). This radii ratio has been suggested to trace quenching and has been shown to depend on the environment in observations \citep{2017MNRAS.464..121S}. In the practice, our prediction could be tested by comparing H$\rm{\alpha}$ with stellar continuum derived sizes.

In summary, in agreement with \citet{Bryant_2018}, we find that most galaxies have angles between 0 and 10 degrees of misalignment, and the fraction of galaxies that are misaligned does not strongly depend on the group mass or the g-i colour. On the other hand, the peak at $\sim180$ in the distribution of the number of galaxies as a function of the PA offset is not clearly obtained in EAGLE . We find a dependence on the fraction of misaligned galaxies with the stellar mass that is not observed in \citet{Bryant_2018}. This could be partially due to the stellar mass ranges adopted here being different to those in \citet{Bryant_2018}, and the small fraction of galaxies with stellar masses $<10^{10}$M$_{\odot}$ used in their study.

Since we find that EAGLE sensibly reproduces misalignments, in what follows, we will use data from this simulation to study the physical drivers that can help us interpret the observations.

\subsection{Relation between misalignments and internal properties}\label{charac}

This section will study the existence of a relationship between the internal properties of galaxies and the misalignment between their stellar and star-forming gas components.

\begin{table*}

\begin{adjustbox}{width=17.8cm}
\begin{tabular}{lllll}

\cline{3-5}
                       &  & \multicolumn{3}{c}{{Number of galaxies}} \\ \hline
\multicolumn{1}{c}{{Population label}} & {Description} & ${z=0.87-1.00
}$ & ${z=0.50-0.62}$ & ${z=0.00-0.10}$  \\ \hline \hline
\multicolumn{1}{l}{Remained misaligned} & \makecell[l]{Galaxies that were and remain misaligned since \\ the previous snapshot.}  &2409 (11$\pm$0.2$\%$)      &1413 (7$\pm$0.2$\%$)    &990 (6$\pm$0.2$\%$)  \\ \hline 
\multicolumn{1}{l}{Remained aligned } &  \makecell[l]{Galaxies that are aligned and remain so \\  in two consecutive snapshots.}  &11375 (56$\pm$0.6$\%$)     &13205 (69$\pm$0.7$\%$)    &12037  (76$\pm$0.9$\%$)   \\ \hline 
\multicolumn{1}{l}{Misaligned} & \makecell[l]{Galaxies that become misaligned between \\ consecutive snapshots.} &2605 (12$\pm$0.2$\%$)     &2159 (11$\pm$0.2$\%$)   &1474 (9$\pm$0.2$\%$)   \\ \hline
\multicolumn{1}{l}{Aligned} & \makecell[l]{Galaxies that were misaligned in the previous snapshot\\ and become aligned in the present one.}  &3886 (19$\pm$0.3$\%$)    &2163 (11$\pm$0.2$\%$)    &1266  (8$\pm$0.2$\%$)  \\ \hline

\end{tabular}
\end{adjustbox}
  \caption{Description of the populations analysed in Section \ref{caract}. The number of galaxies and the percentage of galaxies of each population is tabulated for each redshift range. Errors correspond to Poisson uncertainties but do not consider possible small fluctuations that can take a galaxy just above/below the threshold angle used to classify galaxies as misaligned. This sample is limited to galaxies that meet the selection described in Section \ref{sampleselection} for two consecutive snapshots.}
  \label{tab:tab1}
\end{table*}

\subsubsection{Aligned and misaligned galaxy populations}

First, we make a general characterisation of the aligned and misaligned galaxy populations according to their physical properties. We remind the reader that we measure the angle between the angular momentum vector of the stars and the star-forming gas, considering all particles within an effective radius. In Fig. \ref{histangle} we show the sample distribution of PA offsets at the three redshift ranges. At $z=0.00-0.10$, 20$\pm$0.3$\%$ (3372/16313) galaxies are misaligned, i.e., have PA offset $>30^{\circ}$. At $z=0.50-0.62$, 23$\pm$0.3$\%$ (5139/21927) of the galaxies are misaligned, and at $z=0.87-1.00$ 28$\pm$0.4$\%$ (6348/22433) of the galaxies are misaligned. Misalignment decreases with cosmic time. With the advent of large IFU galaxy surveys at intermediate cosmic times (e.g. the Middle Ages Galaxy Properties with Integral Field Spectroscopy, MAGPI,  \citealt{Foster2021}), it will be possible to measure the misalignment fraction at z>0 and test these predictions.

Fig. \ref{colorssfr} shows the PA angle between the angular momentum vectors of the stars and star-forming gas as a function of halo mass (left panel) and stellar mass (right panel), both coloured by the sSFR at $z=0.0-0.1$. In general, at fixed group mass and stellar mass, the sSFR is higher for galaxies with PA offset lower than 30$^{o}$. We also study this trend until $z=1$, finding that the sSFR decreases over time and with increasing galaxy group mass. Additionally, sSFR decreases as galaxy stellar mass increases at all redshifts in this study, in qualitative agreement with the results of \citet{2007ASSP....3..487B} from MUNICS and FDF observations. On average, misaligned galaxies also become more frequent at higher halo masses and higher stellar masses with time.

\begin{figure}
\begin{center}
\includegraphics[width=7.5cm]{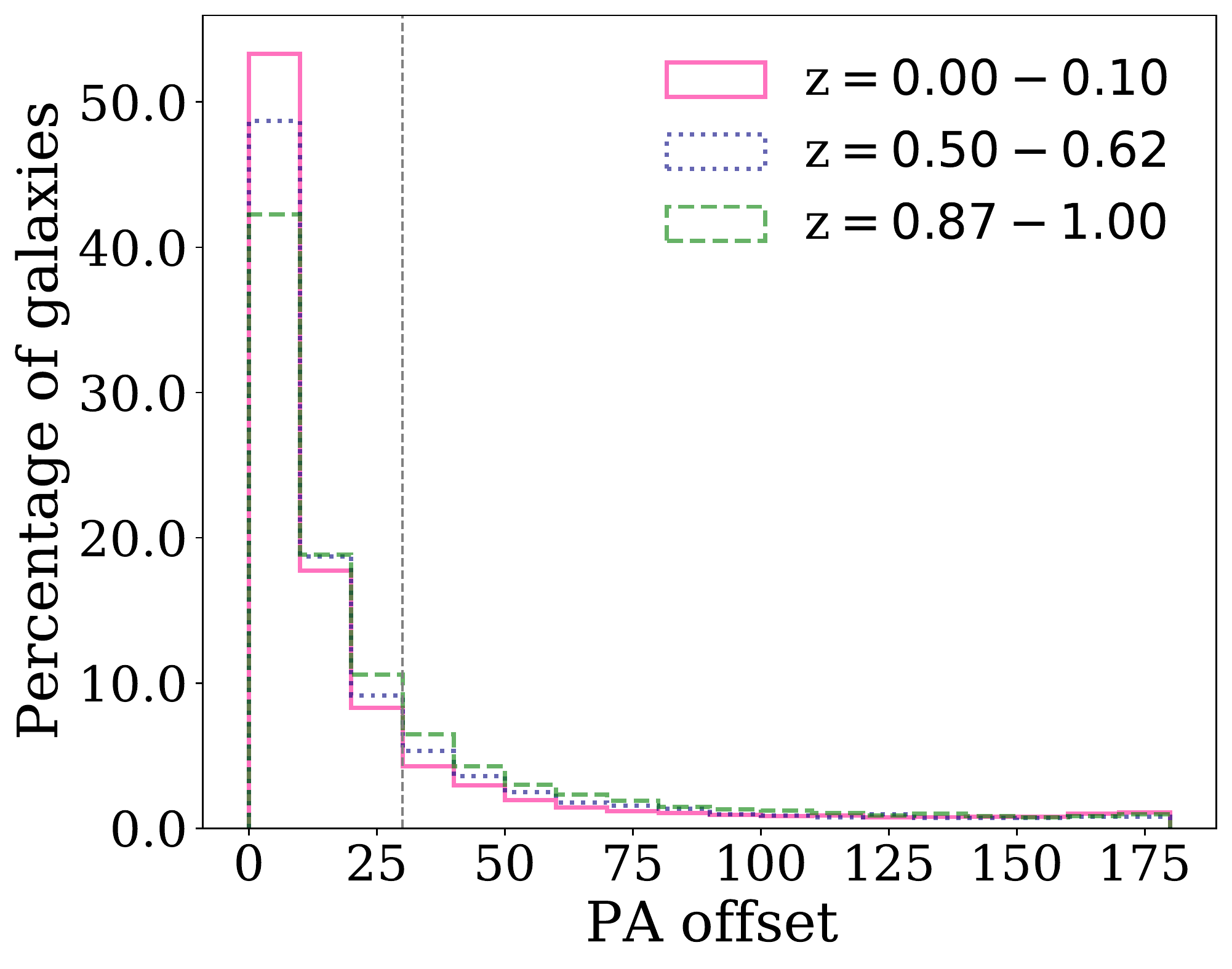}
\caption{Distribution of the angles between the rotation axis of the stars and star-forming gas for galaxies at the three redshift ranges, as labelled. The grey dashed line marks PA offset $= 30^{\circ}$ to separate between aligned and misaligned galaxies. }\label{histangle}
\end{center}
\end{figure}

\begin{figure}
\begin{center}
\includegraphics[width=8cm]{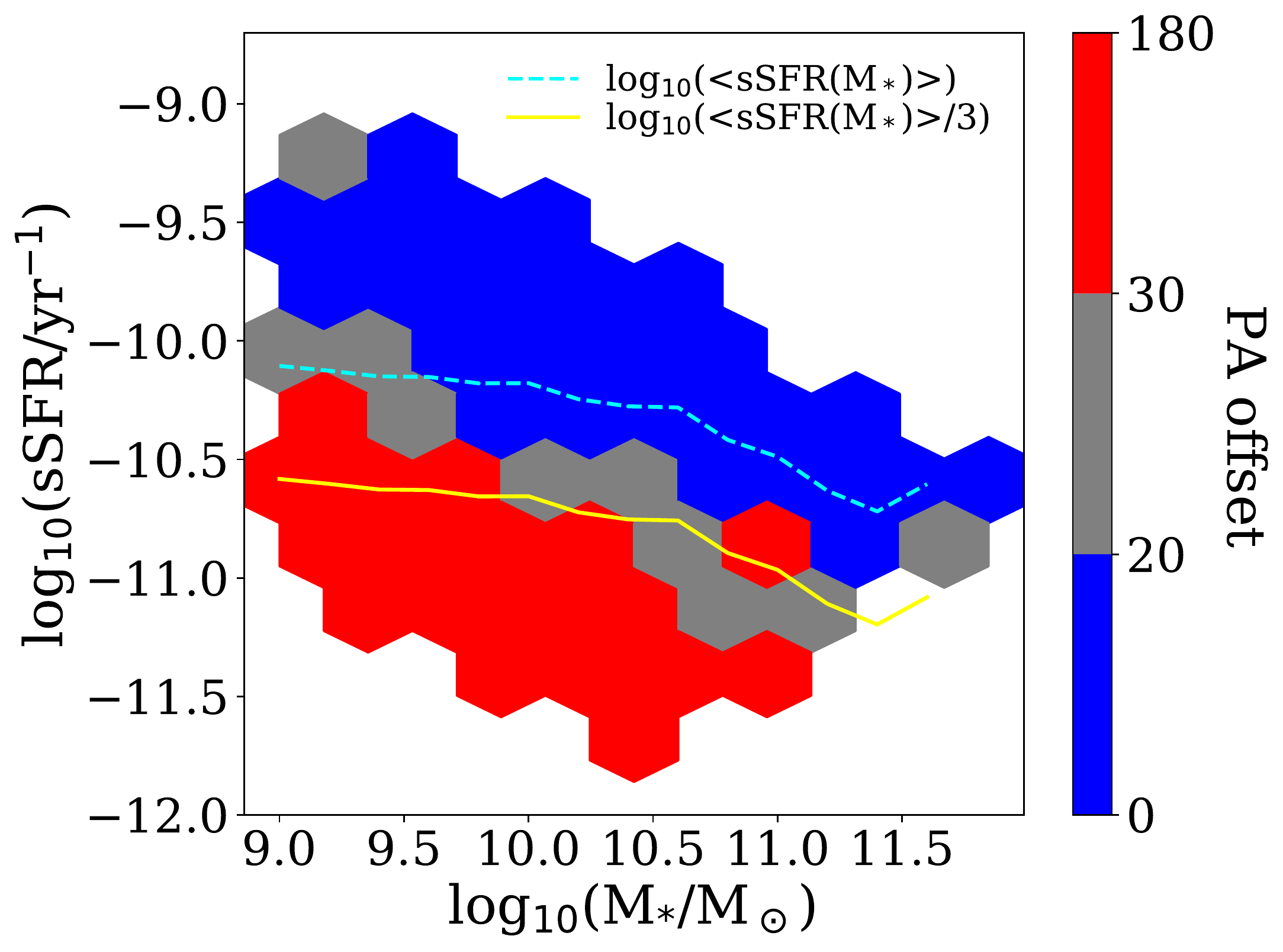}
\caption{Binned distribution of galaxies in the sSFR-M$_*$ plane coloured by the average gas-stellar position angle offset at $z=0.0-0.1$l. The dashed blue line marks the main sequence of star formation and solid yellow line separates the population between blue cloud and red sequence. Only bins with more than five galaxies are shown. }\label{pass-act}
\end{center}
\end{figure}

Since observational studies focus on different morphological types, we separate between red sequence and blue cloud using the sSFR-M$_{*}$ criteria. This method is based on the position of the main sequence of star formation in galaxies, which we quantify with the median sSFR of galaxies with a star formation cutoff defined as log$_{10}$(sSFR$_{\rm{MS}}$/yr$^{-1}$) > $-11+0.5z$ \citep{2015MNRAS.450.4486F}. We calculate the median sSFR of the star-forming main sequence (MS), $<$sSFR(M$_{*})$$>$, using 15 equidistant bins in log$_{10}$ stellar mass in the range of $10^9$ to $10^{12}$ M$_{\odot}$. We then consider as passive/star-forming galaxies those below/above $c$$<$sSFR(M$_{*})$$>$, where $c$ is relatively arbitrary and different values are used in the literature. For example \citet{2015A&A...573A.113B} use $c=1/4$, while $c=1/2$ is used in \citet{2019MNRAS.487.3740W}. In Fig. \ref{pass-act} we adopt $c=1/3$, in between these two suggested values.

In Fig. \ref{pass-act}, we show the sSFR-stellar mass plane, colouring bins by the mean PA offset angle at $z=0.0-0.1$. The zero-point of the main sequence of star formation decreases over time (we study this until $z=1$) simultaneously as more galaxies move towards the passive population. Also, misaligned galaxies mostly belong to the passive population, in agreement with the works of \citet{2011MNRAS.414..968D} using the ATLAS$^{\rm{3D}}$ survey and \citet{Bryant_2018} using SAMI observations. We also obtain this trend using the $\kappa_{\rm{co}}$, $v_{\rm{rot}}/\sigma_{0}$ and $D/T$ thresholds proposed by \citet{2017MNRAS.472L..45C} and \citet{Thob_2019} to separate between ETGs and LTGs  (see Table \ref{perclt}). 

We find that $\sim 24\%$ of EAGLE ETGs are misaligned, consistent with the \citet{2020ApJ...894..106K} Horizon-AGN results (23.7$\pm$0.2\%) but quite different from the SAMI observations (32.7$\pm$6.6\%). This discrepancy could be explained by the fact that SAMI galaxies are classified as ETGs and LTGs by visual inspection and not through the kinematic thresholds calculated in simulations. As an example, the sample of ETGs in simulations could be reduced due to classifying some S0 galaxies as LTGs. On the other hand, we find that $\sim 5\%$ of LTGs in EAGLE are misaligned. This result is comparable with the misalignment fractions of LTGs in Horizon-AGN (6.7$\pm$0.1\%; \citealt{2020ApJ...894..106K}) and SAMI (5.2$\pm$0.7\%; \citealt{Bryant_2018}).

Note that the correlation between the presence of misalignments and the sSFR at fixed stellar mass of Fig.~\ref{pass-act} is much stronger than the weak correlation seen with galaxy optical colour in Fig.~\ref{smygi}. This is not necessarily surprising as it is well known that the correlation between galaxy colour and sSFR saturates as sSFR decreases (e.g. \citealt{Leja2019,Bravo2021}).

\begin{figure*}
\begin{center}
\includegraphics[width=\textwidth]{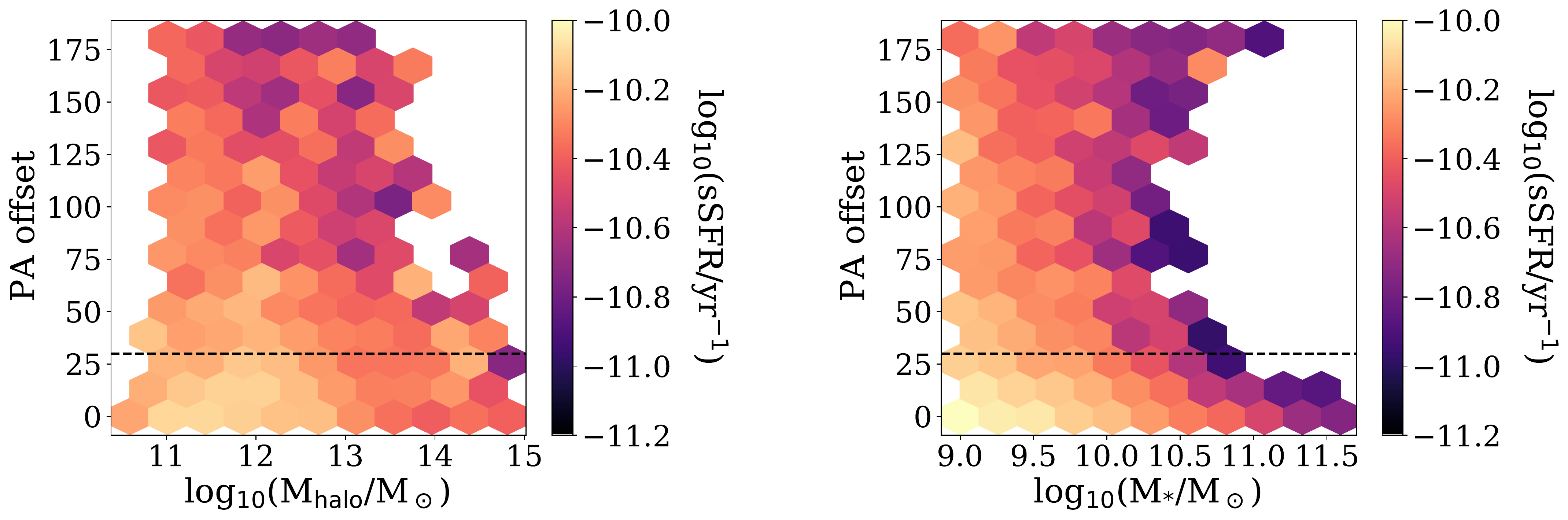}\addtocounter{figure}{+1}
\advance\leftskip-0.18cm
\caption{Binned distribution of the angle between the stellar and star-forming gas angular momentum vectors versus halo mass of the group in which the galaxy resides (left panel) and versus the stellar mass (right panel) coloured by the average Log sSFR, as shown in the colour bar, for galaxies at $z=0.0-0.1$. Only bins with more than five galaxies are shown. The black dashed lines mark a PA offset $= 30^{\circ}$ to separate between aligned and misaligned galaxies.}\label{colorssfr}
\end{center}
\end{figure*}

\subsubsection{Evolution of the alignment state}\label{caract}

In what follows, we divide galaxies into four populations according to Table \ref{tab:tab1}. This was necessary in order to distinguish between internal properties that cause galaxies to stay aligned/misaligned or become aligned/misaligned in consecutive snapshots. To test the sensitivity of our results to the exact PA offset threshold we use to classify galaxies as misaligned, we choose higher values and find the trends remain qualitatively the same and our conclusions hold.

\begin{figure*}
\begin{center}
\includegraphics[width=0.85\textwidth]{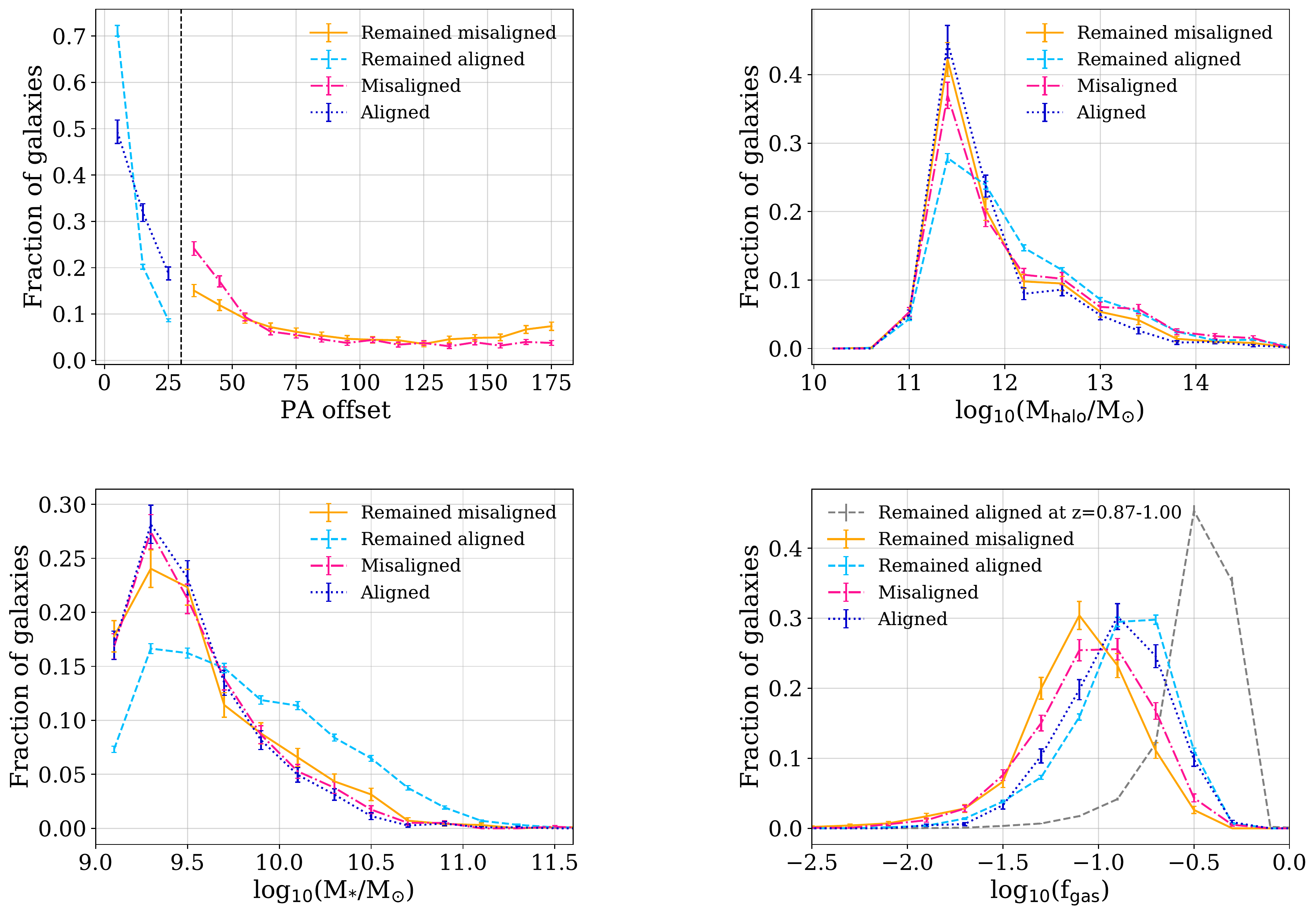}\\
\vspace{0.5cm}
\includegraphics[width=0.85\textwidth]{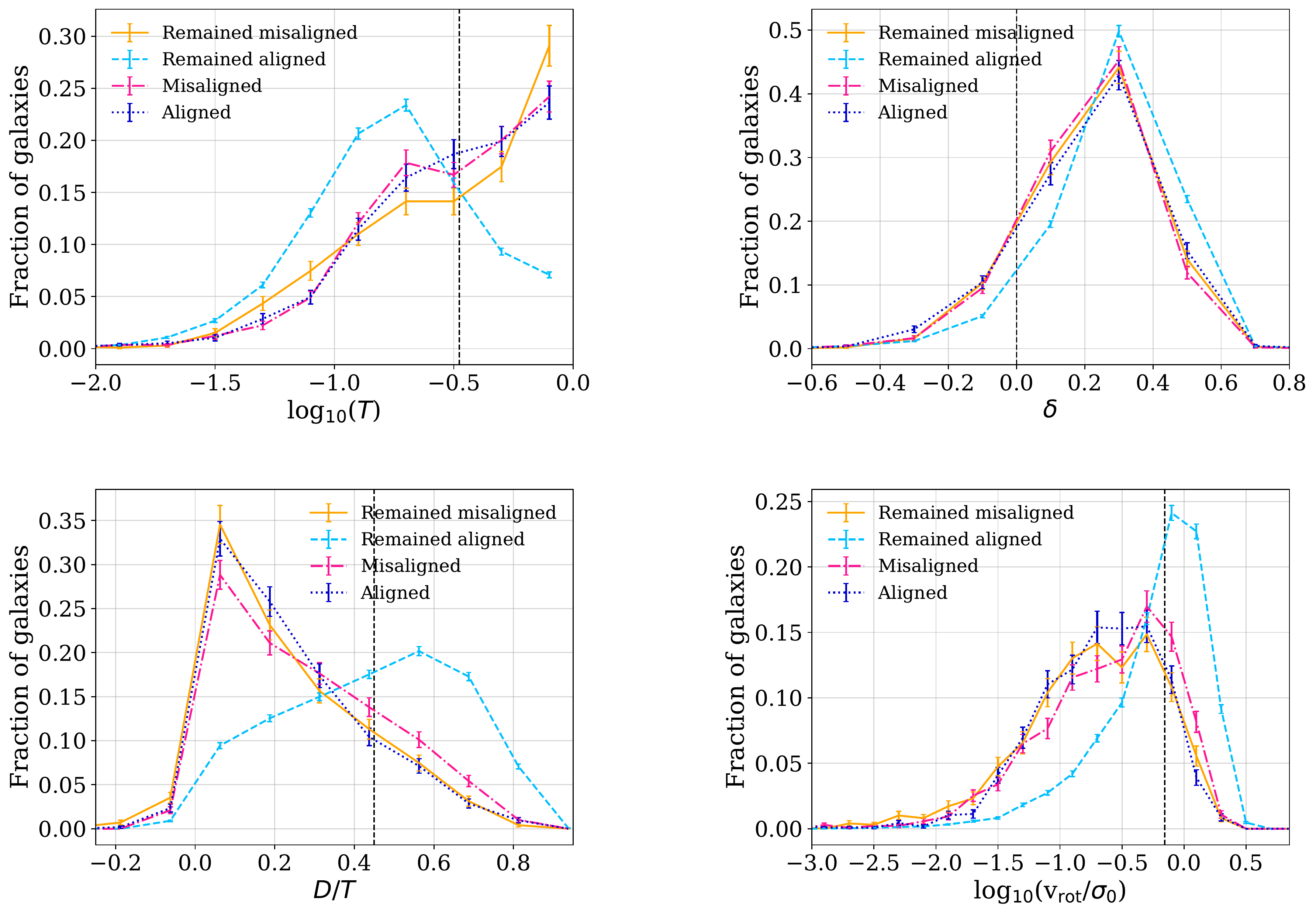}
\advance\leftskip0.2cm
\caption{Each panel shows the distribution of galaxy properties normalized to each population described in Table \ref{tab:tab1} at $z=0.00-0.10$. From top left and counterclockwise, galaxy properties are: gas-stellar position angle offset, stellar mass, triaxiality parameter (the black dashed line marks $T$=1/3 to separate between oblate and prolate systems), the disc-to-total stellar mass ratio (the black dashed line signalize $D/T = 0.45$ to divide between star-forming and passive galaxies), the ratio between the stellar rotational velocity and the velocity dispersion (the black dashed line marks $v_{\rm{rot}}/\sigma_{0} = 0.7$ to separate between star-forming and passive galaxies), stellar velocity anisotropy parameter, gas fraction, and halo mass. The fractions in each bin sum to 1. Errors correspond to Poisson uncertainties.}\label{propiedadesall}
\end{center}
\end{figure*}

\begin{figure*}
\begin{center}
\includegraphics[width=\textwidth]{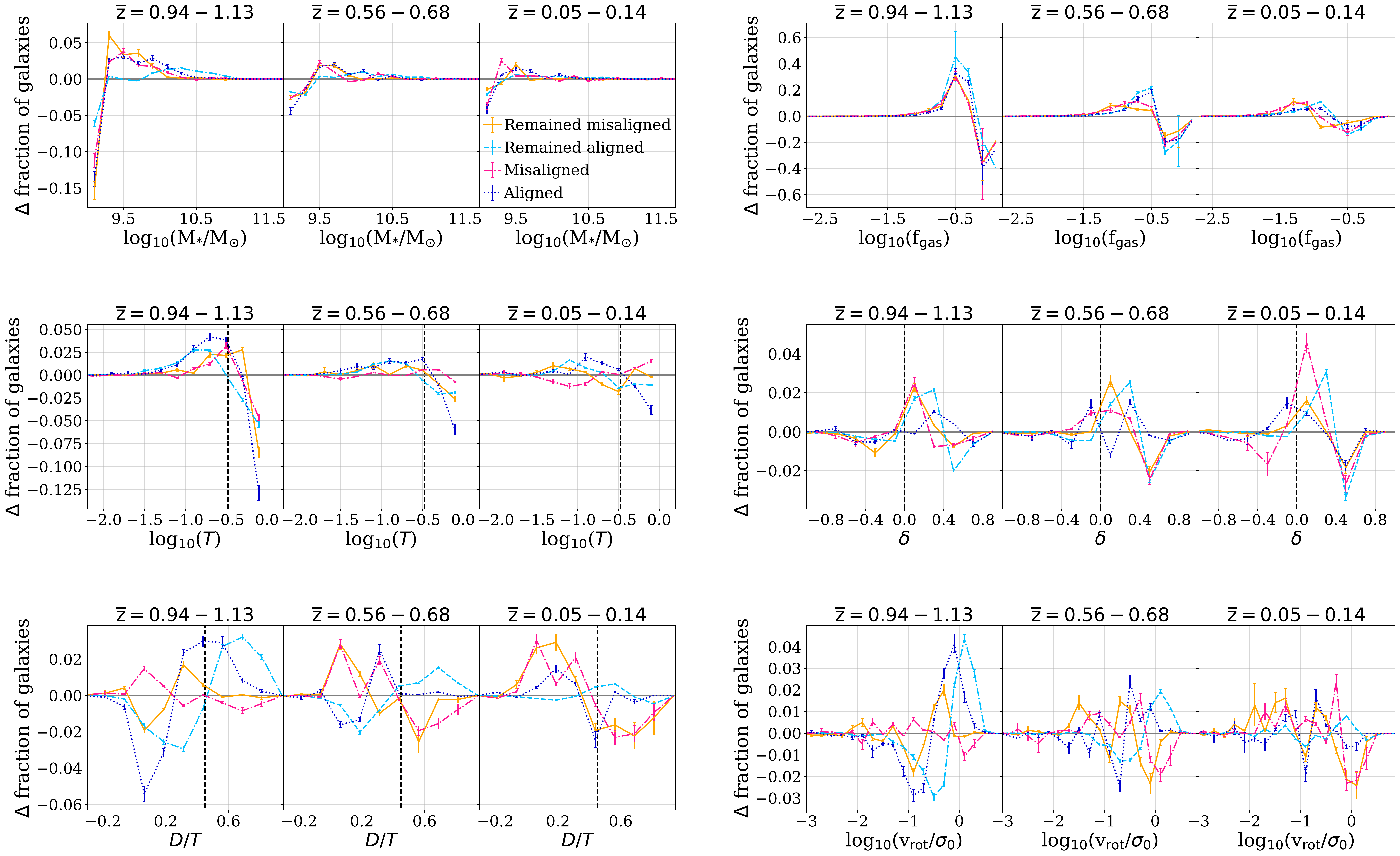}
\caption{Each chart shows the difference in the abundance of galaxies in each bin between two consecutive snapshots. We show this for all the properties studied in Fig. \ref{propiedadesall}. A positive/negative value of `$\Delta$ fraction of galaxies' means that the number of galaxies increases/decreases in each galaxy property bin, between two consecutive snapshots. Errors correspond to Poisson uncertainties.}\label{diffe}
\end{center}
\end{figure*}

Fig. \ref{propiedadesall} shows the distribution of different galaxy properties for each of the four populations of Table \ref{tab:tab1} at $z=0.00-0.10$. We will describe the results for each property, starting from the panel at the top left and continuing to the right and from top to bottom.

Regarding the distribution of the PA offset, galaxies that remain aligned have a more prominent peak at offsets under 10$^{\circ}$, while the population of `aligned' galaxies have a less prominent peak at PA offset $\sim$ 0°, with a higher fraction of galaxies with 10°$<$PA$<$30°. The population of `remained misaligned' galaxies have the flattest distribution of PA offset and a clearer secondary peak at $\sim$180° compared to those that become misaligned. Also, galaxies that remain aligned reside on average in more massive halos than the rest of the populations. 

Galaxies in the `remained aligned' group have higher stellar masses than the other populations. Nevertheless, most `remained aligned' galaxies seem to be low-mass galaxies (see Table \ref{anex1} for specific percentages). This trend remains the same over time until $z=1$. We also find that stellar mass increases are more significant at higher redshift (see Fig. \ref{diffe}), in agreement with the overall history of stellar mass growth inferred from observations \citep{2018MNRAS.475.2891D}, which shows that stellar mass increases steeply in the early Universe to become slower towards z < 0.5.

In agreement with observations, misaligned galaxies or remain misaligned have a smaller fraction of star-forming gas than other galaxies. This difference is greater at low redshift.

The decrease in the gas fraction is more notorious at high redshift, being greater for galaxies that remain aligned, which is not necessarily surprising, as galaxies that are misaligned are already gas-poor (see Fig. \ref{diffe}). The latter has a cascade effect, as fewer stars form because of the gas-poorness, leading to fewer outflows and hence fewer chances of losing the existing gas in the galaxy. Table \ref{anex2} tabulates the percentage of galaxies that are below $f_{\rm{gas}}$ = 0.1. It is important to mention that this limit does not represent a division between gas-poor and gas-rich galaxies because, in order to determine this, it would be necessary to make a comparison with a linear fit of the `gas-rich sequence', the gas fraction vs stellar mass relation of star-forming galaxies. Nevertheless, it allows us to make a qualitative comparison that we can complement with the stellar mass distribution.

In other hydrodynamic simulations such as Horizon-AGN and Illustris, it has also been found that galaxies displaying misalignments are the gas poorest. Conversely, \citet{2015MNRAS.448.1271L}, using the \textsc{galform} semi-analytic model of galaxy formation, found that ETGs with low cold gas fractions are more likely to display aligned cold gas and stellar components compared to ETGs that have higher gas fractions. However, their model did not consider any relaxation of the gas disc towards the stellar component due to torques, while hydrodynamic simulations naturally take into account relaxation processes that affect the distribution of misalignments. This difference could be significant if we study both ETGs and LTGs since, in the latter case, torques would be stronger since the gas has a clear disc to which to relax.

Galaxies that remain aligned are mostly oblate, while other populations are mostly triaxial on prolate (see Table \ref{anex3}). We find that at $z= 0.87-1.00$ all populations are mostly triaxial on prolate, but the largest fraction of oblate is found for galaxies that remain aligned. Changes in triaxiality become stronger as the redshift increases (see Fig. \ref{diffe}), which can be interpreted as morphology changing more dramatically at high redshift.

Moreover, galaxies that remain aligned have the highest values of velocity anisotropy, i.e., they are flatter than those galaxies that become misaligned or remained misaligned. However, there is not much difference between populations and in most galaxies, the velocity dispersion is primarily contributed by disordered motion in the disc plane, i.e., have positive values $\delta$ (see Table \ref{anexani}).

The population of `remained aligned' galaxies also has the highest values of disc-to-total ratio (i.e., have a higher fraction of stars that are rotationally supported), while galaxies that remain misaligned have the lowest values. In Table \ref{anex33} we show the percentage of galaxies that are below the $D/T = 0.45$ limit, suggested by \citet{Thob_2019}, that best separates between LTGs and ETGs. According to this, the most disc-dominated galaxies are those that remained aligned in time.

Galaxies that remain aligned have the highest values of $v_{\rm{rot}}/\sigma_{0}$, i.e., are more supported by rotation than the other populations and tend to have disk-shaped structures due to the highly aligned motion of stars. Galaxies that remain misaligned have the lowest $v_{\rm{rot}}/\sigma_{0}$, i.e., dispersion is more significant, and they tend to have spheroidal shape structures. This result is consistent with observations, with the Horizon-AGN results from \citet{2020ApJ...894..106K} and with the IllustrisTNG100 results from \citet{Duckworth2020a}. The elliptical shape of the galaxy stellar mass distribution affects the time it takes for the gas disk to torque towards the stellar disk.

In Table \ref{taab} we show the percentage of early-type galaxies, at each redshift range, predicted by the $D/T$ and $v_{\rm{rot}}/\sigma_{0}$ thresholds suggested by \citet{Thob_2019}. We find that at $z = 0.87 - 1.00$ the difference between predicted percentages is $\sim6\%$, at $z = 0.50 - 0.62$ is $\sim8\%$ and at $z=0.00-0.10$ is $\sim10\%$. They overall agree well.

Fig. \ref{propiedadesall} shows that there are important differences between the four galaxy samples of Table \ref{tab:tab1}. Overall, the galaxy population of galaxies that `remain misaligned' are the gas-poorest, are the most dispersion dominated, have the smallest disc contribution, and are the most prolate. On the other hand, galaxies that remain aligned are the gas-richest, most rotation dominated, have the most significant disc contributions, and are the most oblate. In general, we find that galaxies that display misalignments, in either the remain misaligned, misaligned or become aligned populations, are significantly different from the population that does not display any misalignments (remain aligned). This strongly suggests that misalignments are present in a particular galaxy population.

\begin{table}
\centering
\begin{adjustbox}{width=8.49cm}
\begin{tabular}{cccc}

\cline{2-4}
                        & \multicolumn{3}{c}{{Percentage of early-type galaxies predicted (\%)}} \\ \hline
\multicolumn{1}{c}{{Threshold}} & ${z=0.87-1.00
}$ & ${z=0.50-0.62}$ & ${z=0.00-0.10}$  \\ \hline \hline

\multicolumn{1}{l}{$D/T$ = 0.45}  &  59 $\pm$ 0.7     &  53 $\pm$ 0.7  & 54 $\pm$ 0.7  \\ \hline
\multicolumn{1}{l}{$v_{\rm{rot}}/\sigma_{0}$ = 0.7} & 63 $\pm$ 0.7      &  55 $\pm$ 0.7     &    54 $\pm$ 0.6   \\ \hline

\end{tabular}
\end{adjustbox}
  \caption{Percentage of early-type galaxies in the sample defined in Section \ref{caract},  using the thresholds suggested by \citet{Thob_2019}. Errors correspond to Poisson uncertainties.}
  \label{taab}
\end{table}

We now move to assess the influence of external parameters of galaxies that relate to the environment they live in. This is done to establish whether misalignments care more about the internal properties of galaxies or their environment. The tidal strength parameter Q is an estimation of the total gravitational interaction strength that the neighbours produce on a galaxy with respect to its internal binding forces \citep{2015A&A...578A.110A}. To measure Q, we consider as neighbours all galaxies within the same halo with a stellar mass higher than $10^8$M$_{\odot}$.

\begin{figure*}
\begin{center}
\includegraphics[width=0.9\textwidth]{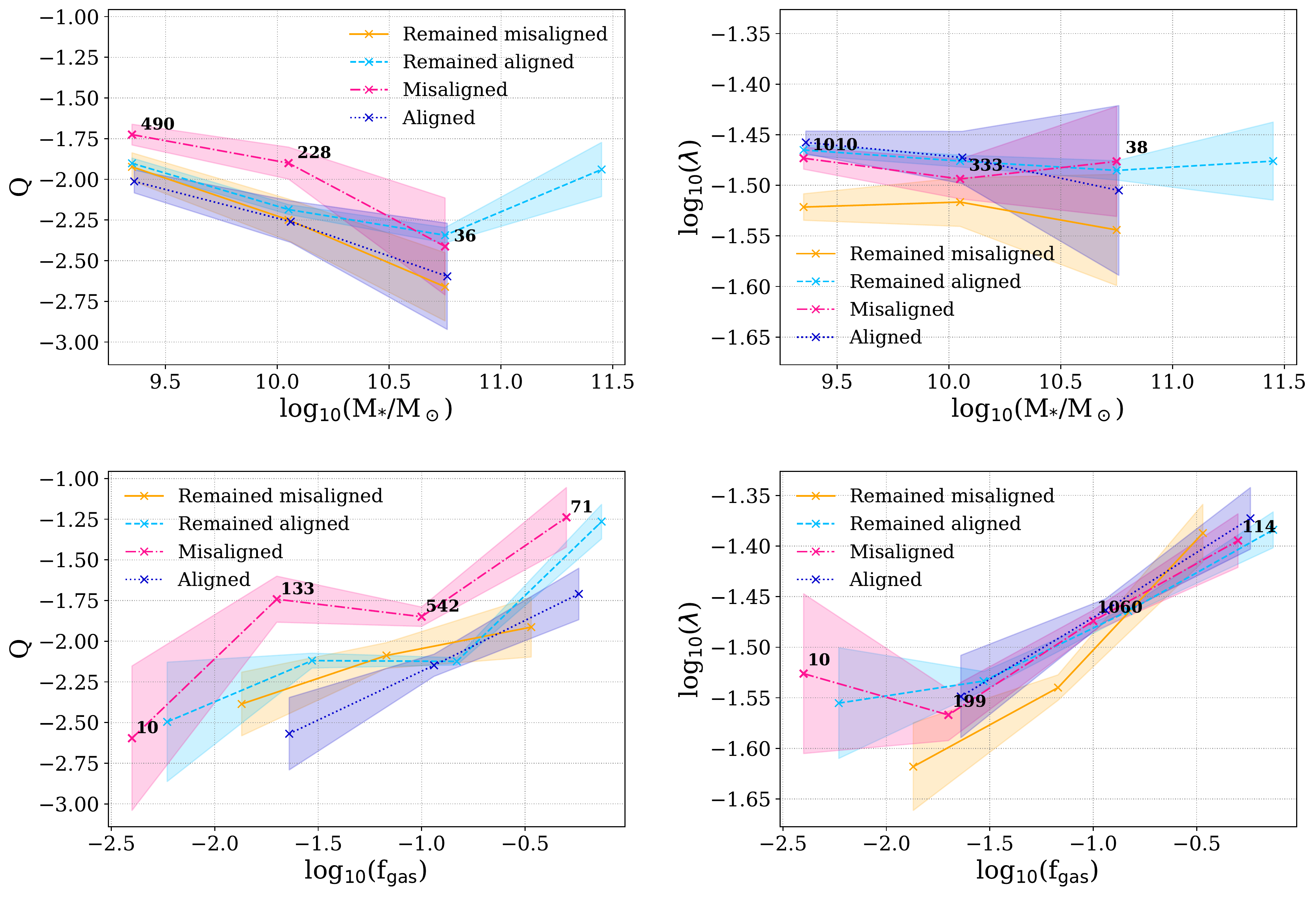}
\advance\leftskip-0.85cm
    \caption{Left panels show the average tidal strength parameter Q in bins of stellar mass (top panel) and bins of gas fraction (bottom panel) for each population described in Table \ref{tab:tab1} at $z=0.00-0.10$. Same for right panels but for the halo spin parameter $\lambda$. The number of galaxies considered in each bin for the `misaligned' population is shown, and the shaded area represents the Poisson error.
    }\label{Qylambda} 
\end{center}
\end{figure*}

\begin{figure}
\begin{center}
\includegraphics[width=8cm]{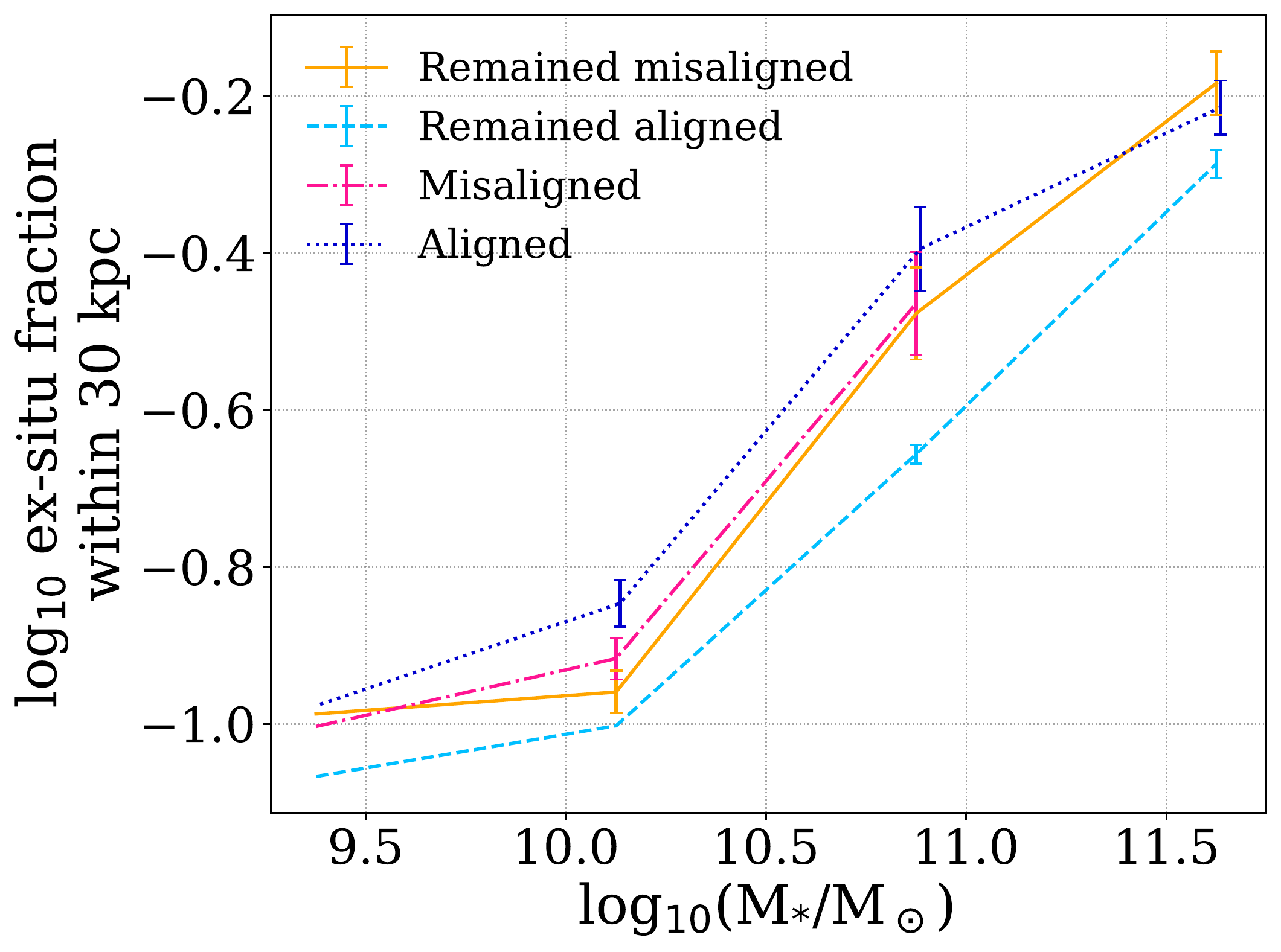}
\caption{Distribution of ex-situ stellar fraction in stellar mass bins for each population described in Table \ref{tab:tab1} at $z=0.0-0.1$. Only bins with more than five galaxies are shown. Error bars are Poisson errors.}\label{ex-situ}
\end{center}
\end{figure}

Fig. \ref{Qylambda} shows the tidal strength parameter Q as a function of stellar mass and gas fraction. In the bins with the best statistics, the galaxies that become misaligned between two consecutive snapshots are generally more affected by neighbours than the other populations. This suggests that the physical processes related to misalignment could be due to interactions with neighbouring galaxies, but note this does not necessarily indicate galaxy mergers as the cause. To obtain statistically significant results on the relationship between the change of the tidal parameter over time and the misalignments, a larger simulation is needed but with a resolution comparable to that of EAGLE.

Fig. \ref{Qylambda} also shows the spin parameter of the host dark matter halo, $\lambda$. For `aligned,' `misaligned', and `remained aligned' populations, $\lambda$ is  similar at fixed stellar mass and fixed gas fraction. The halo spin parameter is different (lower) for galaxies that remain misaligned compared to other populations, suggesting that there must be special conditions for galaxies to remain misaligned for an extended period of time. Beyond the tidal field strength and halo's spin, we also explored possible correlations with halo mass, the volume and surface density of galaxy neighbours, central/satellite definition, and halo concentration, finding none. For example, we explored (i) the neighbour density (the volume density to the $3^{\rm{rd}}$, $7^{\rm{ th}}$ and $10^{\rm{th}}$ nearest neighbour) finding no difference between aligned and misaligned galaxies; (ii) the frequency of minor mergers (explored using the high time cadence of the EAGLE snipshots), defined as those with a stellar mass ratio $<0.1$, finding no difference between aligned and misaligned galaxies; (iii) the gas accretion rate to the galaxies as defined in \citet{2020MNRAS.495.2827C} and we found a weak trend of misaligned galaxies having lower accretion rates, which is expected due to the correlation with sSFR; (iv) the gas and stars being stripped from the galaxies, finding no difference between aligned and misaligned galaxies.

We explore the effects of interactions by studying the cumulative effect via the fraction of the stellar mass of galaxies that formed ex-situ. We considered a particle as ex-situ if the subhalo to which it belonged at the snapshot prior to star formation is not in the main branch of the final galaxy, as described in \citet{2020MNRAS.497...81D}.

Fig.~\ref{ex-situ} shows the distribution of ex-situ stellar fraction within 30 kpcs at fixed galaxy stellar mass. The `remained aligned' galaxies have the lowest fraction of stellar mass accreted, showing a lighter cumulative effect of mergers and interactions in comparison with populations that experienced a change in their state of alignment or remained misaligned. This trend is consistent with the result of \citet{2020ApJ...894..106K}, who found a significant contribution of mergers and interaction with nearby galaxies to misalignments (of about 57$\%$) in Horizon-AGN.

From all the relations studied in this section, it appears that despite the presence of some correlations between the environment (using a large variety of definitions as listed above) and the incidence of misalignments, the internal properties of galaxies are better indicators of whether misalignments are expected. A way of interpreting this is that the environment triggers the presence of misaligned SF gas (e.g., via interactions, mergers, or sudden accretion), but the internal properties of galaxies determine whether that SF gas quickly aligns (which can happen even before the gas is properly accreted onto the galaxy) or remains misaligned for long enough as to be seen in between simulation snapshots (which are tracing timescales of about few 100 Myr).

\subsection{Relation between misalignments and physical processes}\label{proc}

In what follows, we study whether there is a relationship between the physical processes described in Section \ref{sec:ext} and the kinematic misalignment. First, we analyse two study cases in which the stellar and gas components become misaligned, and second, we perform a general statistical study.

\begin{figure*}
    \centering
    \subfigure{\includegraphics[width=8cm]{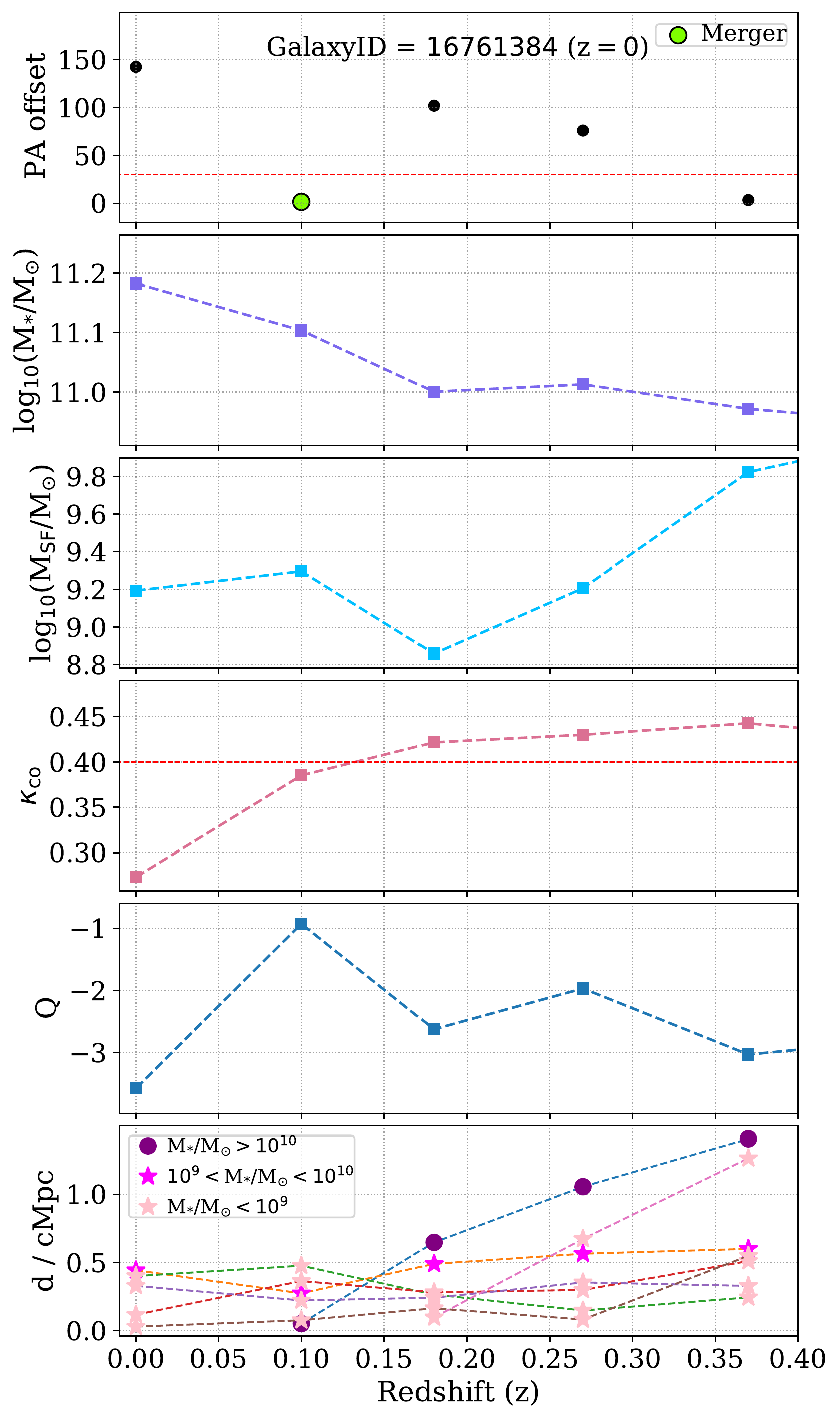}
    }
    \subfigure{\includegraphics[width=8.15cm]{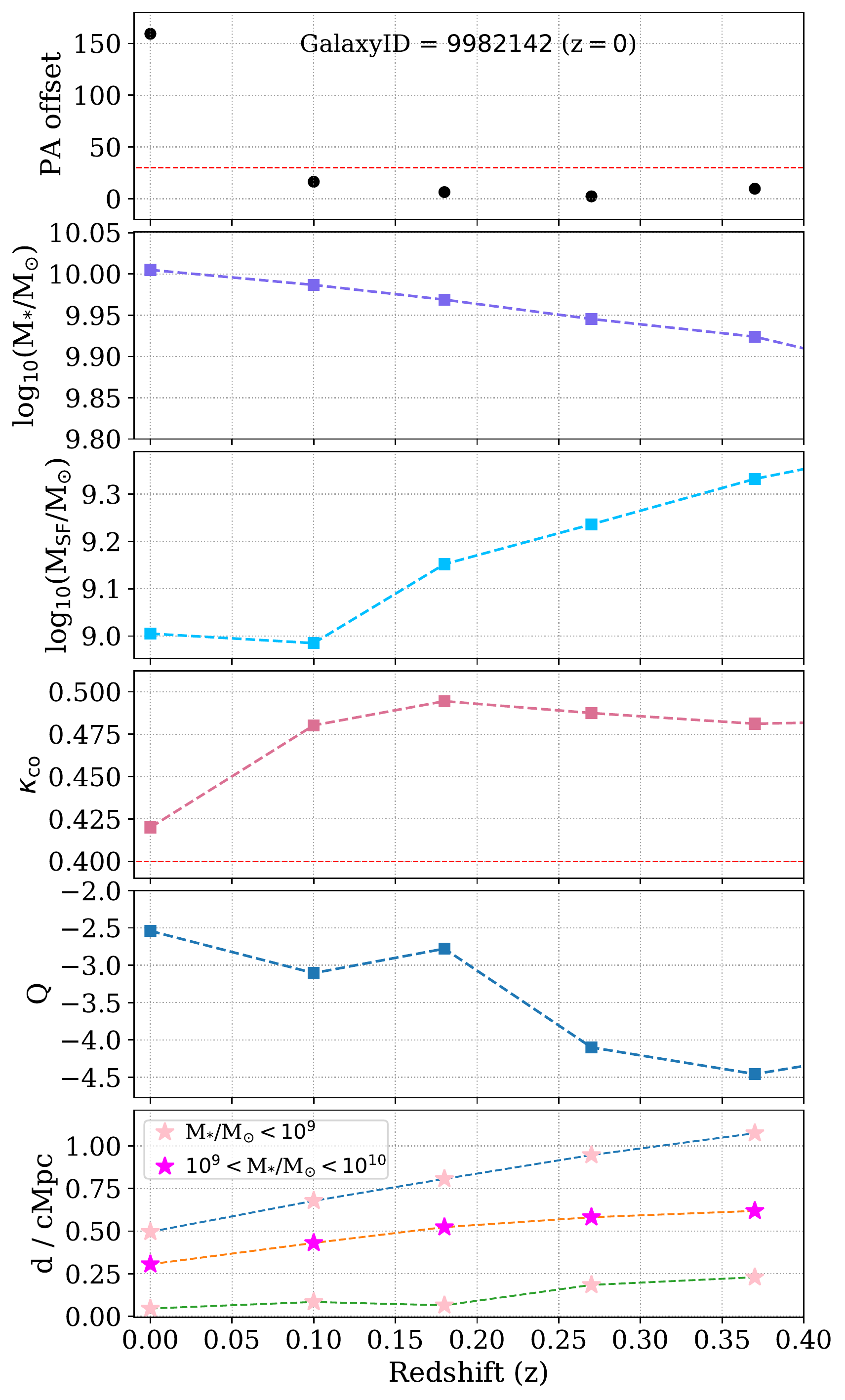}
    }
    \caption{From top to bottom, we show the evolution through redshift for two galaxies of PA offset between the rotational axes of the stars and star-forming gas (the red dashed line marks $30^{\circ}$), stellar mass, star-forming gas mass, stellar co-rotating kinetic energy fraction (the red dashed line marks $\kappa_{\rm{co}}= 0.4$), tidal strength parameter Q, and the distance with neighboring galaxies when the misalignment occurs (closer than 0.7 cMpc), colored by mass ranges. In the upper panel on the left, the green dot indicates where the merger occurred.}
    \label{fig:history}
\end{figure*}

In Fig. \ref{fig:history} we show the PA offset, stellar mass, star-forming gas mass,  stellar co-rotating kinetic energy fraction, tidal strength parameter, and the distance with neighbouring galaxies when the misalignment occurs for two galaxies from $z=0.37$ to $z=0.00$. In both cases, stellar and star-forming gas components become misaligned from $z=0.1$ to $z=0.0$.

In the first study case (left side of Fig. \ref{fig:history}, hereafter `Galaxy 1'), the misalignment relates to a galaxy merger with a merger ratio of 0.21 (marked in green). After this merger, the galaxy stellar mass increases by 20$\%$ and the star-forming gas mass decreases by 21$\%$. Also, according to the \cite{2017MNRAS.472L..45C} threshold, the value of $\kappa_{\rm{co}}$ indicates that there is a change in the morphology of the galaxy, moving further into the early-type population after the merger. Interestingly, this example galaxy becomes an elliptical (based on its value of $\kappa_{\rm{co}}$) {\it before} the galaxy displays a misalignment in the gas component, which fits the average picture we get for all galaxies in EAGLE that display misalignments, as discussed below.

The value of the tidal strength is higher when the galaxy merger occurs, as expected, and the event can also be seen in the last panel, where the galaxy that merges with the main one is marked with purple dots. As discussed in the introduction, mergers have been widely identified as a possible cause of misalignments. However, we can find cases such as `Galaxy 2' (right side of Fig. \ref{fig:history}), where the misalignment is not related to a merger or to an abrupt change in stellar mass or star-forming gas mass (stellar and star-forming gas mass increases only by 4$\%$). However, in Galaxy 2, the misalignment is related to a stronger interaction with neighbouring galaxies. This highlights the complexity of the problem in hand and the need to carefully analyse the evolution of individual galaxies as examples to help clarify how a part of the galaxy population may become misaligned.

\begin{figure*}
\begin{center}
\includegraphics[width=\textwidth]{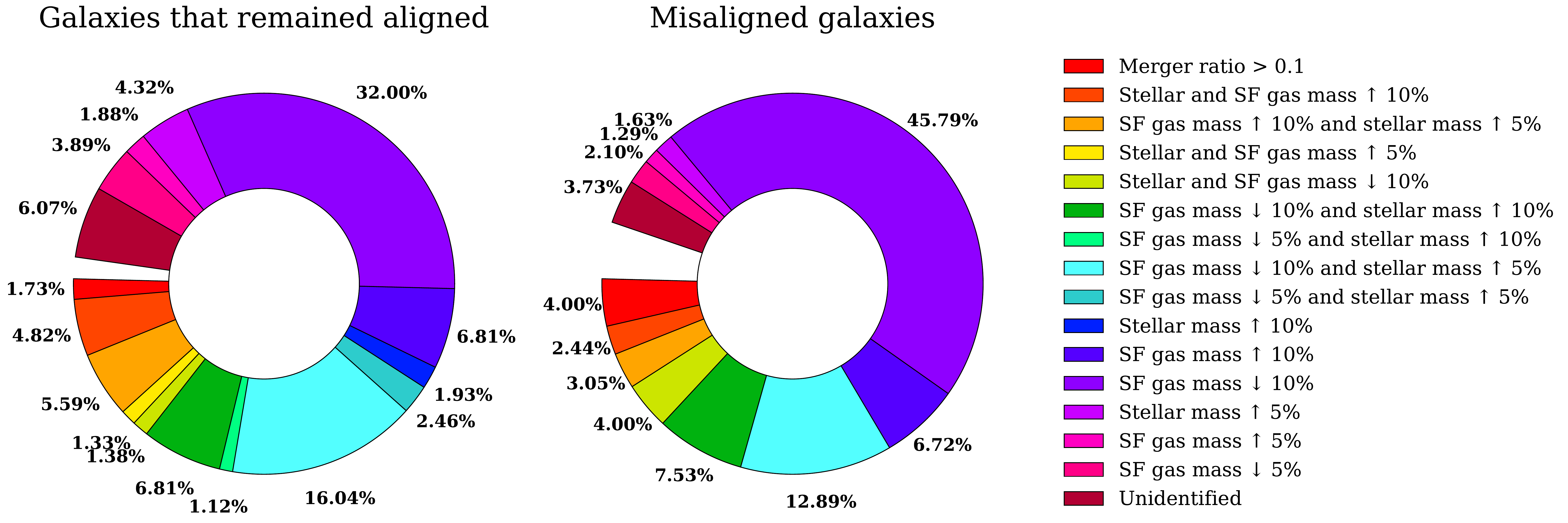}
    \caption{Each pie chart shows the fraction of galaxies of each population (as labelled) at z=0.0-0.1 that suffered a certain event in the previous snapshot. $\uparrow$ $5\%$ and $\uparrow$ $10\%$ mean increases between 5 and 10$\%$ and $\geq$ 10$\%$ , respectively. In these pie charts, we only show events that occurred in more than 1$\%$ of the galaxies of each population.}\label{pie} 
\end{center}
\end{figure*}

\begin{figure}
\begin{center}
\includegraphics[width=8.1cm]{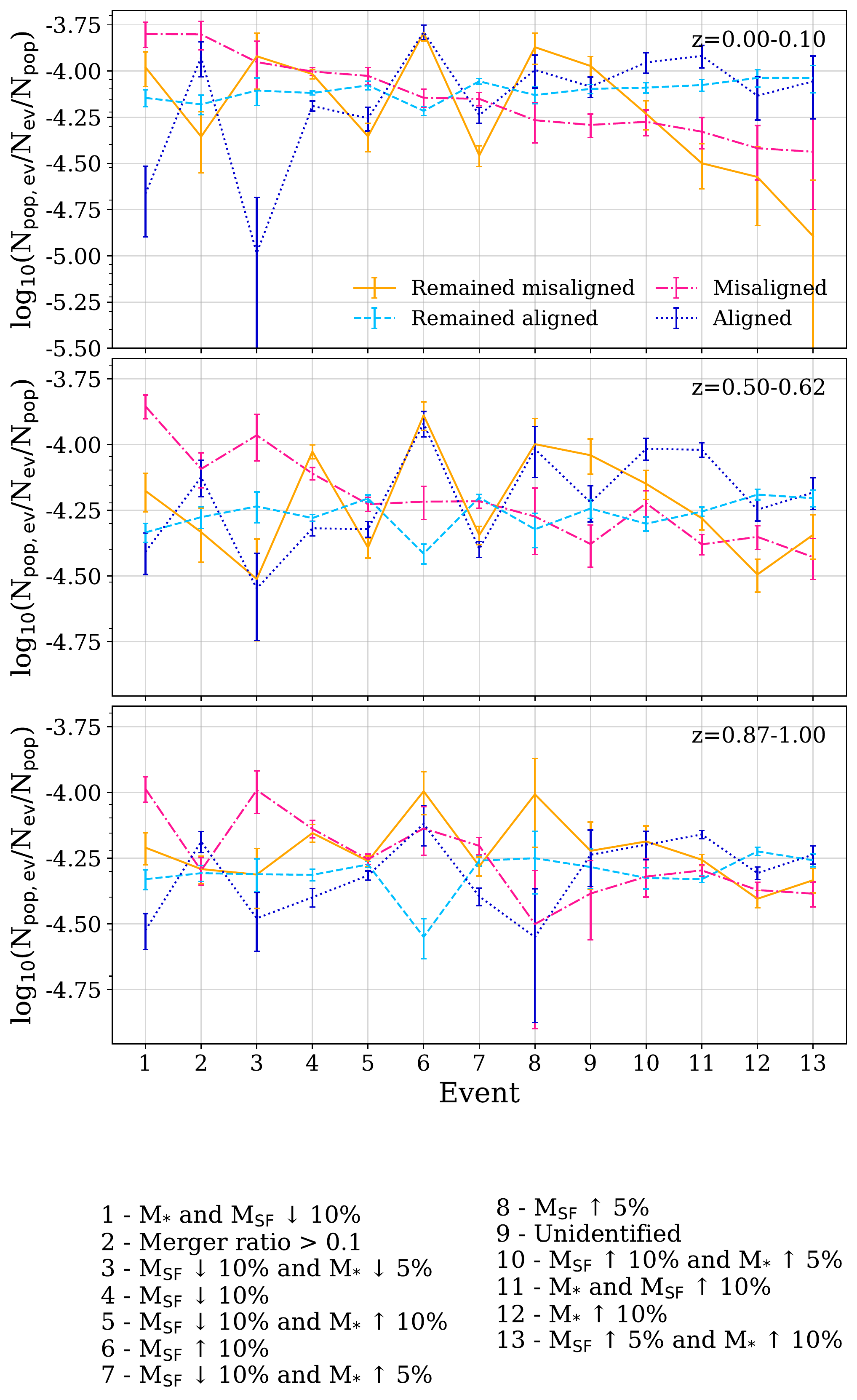}
\advance\leftskip-0.85cm
    \caption{Number of galaxies of each population associated with an event (N$_{\rm{pop,ev}}$), over the total number of galaxies that experienced that event (N$_{\rm{ev}}$), and over the total number of galaxies in that population (N$_{\rm{pop}}$), at three redshift ranges (as labelled on each panel). The integers on the x-axis correspond to each event listed below the figure. Error bars are Poisson errors.}\label{procesos1} 
\end{center}
\end{figure}

Considering mergers and all possible combinations of increase/decrease of stellar/SF gas mass (not related to mergers), we adopt 25 events to classify galaxies within our sample (see the `Event' column in Table \ref{tablapie}). Only a small percentage of the galaxies are not associated with any of these events: 6$\pm$0.2$\%$ at $z=0.00-0.10$, 2$\pm$0.1$\%$ at $z=0.50-0.62$, and 0.5$\pm$0.1$\%$ at $z=0.87-1.00$. The event related to the misalignment of these galaxies was labelled as `unidentified'.

First, we calculate the percentage of misaligned galaxies associated with each event and compare them with the percentage of galaxies that remained aligned. In other words, 100$\%$ corresponds to the total of galaxies that become misaligned between two consecutive snapshots, and we classified them according to which process occurred in the snapshot previous to the misalignment (analogous for `remained aligned'). Fig. \ref{pie} shows pie charts that represent the percentage of aligned and misaligned galaxies at $z=0.0-0.1$ that experienced each event (in Table \ref{tablapie} we tabulated this data including errors for all events). 

The event most associated with both populations is the abrupt decrease in star-forming gas by more than 10$\%$, which simply reflects the average decrease of the SF gas mass in galaxies in EAGLE with time \citep{2015MNRAS.448.1271L}. However, the misaligned population shows a $\approx10\%$ higher preference for this event than the remained aligned one. It is important to highlight that this gas mass loss is not necessarily related to some environmental effect happening predominantly in galaxies that display misalignments. As discussed in $\S$~\ref{caract}, we explored explicitly the gas and stellar mass being stripped from galaxies, finding no significant difference between aligned and misaligned galaxies. Hence, other mechanisms may be responsible for this decrease, such as overall lower gas accretion rates and outflows.

The second most associated event with both populations is the decrease in star-forming gas by more than 10$\%$ accompanied by an increase of stellar mass between 5 and 10$\%$. This event is $\approx3\%$ more likely to be related to galaxies that remained aligned. This simply shows the continuing process of star formation in galaxies, where the decrease in SF gas is related to some of it being transformed into stars and some being expelled from the galaxy via outflows. Remained aligned galaxies are more likely than misaligned galaxies to be in this state.

The percentage of occurrence of all other events is lower than 10$\%$ in both populations. Also, it is worth mentioning that the number of galaxies in which none of the events occurred (classified as `unidentified') is higher for `remained aligned' galaxies, being less than 5$\%$ for misaligned galaxies.

Curiously, the percentage of misaligned galaxies associated with a merger is lower than 5$\%$, but it is still higher than for the `remained aligned' population. This percentage is similar to the one reported in \citet{2015MNRAS.448.1271L} using semi-analytic models, who found that about 6$\%$ of misaligned galaxies come from mergers. We also test the fraction of misalignments associated with very minor mergers (merger ratio < 0.1). We find that, at $z=0.00-0.10$, only 1$\pm$0.2$\%$ of the misaligned galaxies had a very minor merger, 2$\pm$0.3$\%$ at $z=0.50-0.62$, and 2$\pm$0.3$\%$ at $z=0.87-1.00$. Therefore, we discard the possibility of missing an important contribution from very minor mergers in the statistics. The time span between snapshots can be large (up to 1 Gyr), which means we could miss galaxy mergers. To rule out this issue, we analysed the EAGLE snipshots. We found that we were not missing galaxy mergers in our galaxy sample, and hence our estimate of the mergers associated with misalignments is complete. \textcolor{red}{}

This analysis gives us an idea of the distribution of events in the `remained aligned' population compared to the `misaligned' one. We conclude that the event that generates the greatest difference between both populations is the abrupt decrease in star-forming gas by more than 10 $\%$; however, this event is also the most common. To better isolate which processes are preferred by the different galaxy populations, we study what happens after each event, comparing the percentage of galaxies that become misaligned with the other populations.

We calculate the number of galaxies in a population associated with an event, normalised by the total number of galaxies that experienced that event and by the total number of galaxies in that population. In Fig. \ref{procesos1} we show the logarithm of this fraction for the processes that were more statistically related to misalignments at different redshift ranges. 

Misalignments are most related to the abrupt and simultaneous loss of stellar and SF gas mass. Mergers are in second place at $z=0.0-0.1$ but lose importance at higher redshift. The loss of stellar mass suggests interactions with the environment or other galaxies, which is also supported by our previous results on the tidal field. All the other events seem to happen in similar frequency in all the populations.
We also highlight that these events that lead to misalignments are not responsible for the morphological transformation of galaxies into ellipticals in EAGLE. We recall that most of the misalignments in EAGLE happen in elliptical galaxies. However, we find that $\approx 80$\% of the elliptical galaxies that display misalignments, became ellipticals $>1$~Gyr before displaying misalignments, and in $\approx 50$\% of the cases this happened $>2$~Gyr before. Hence, the processes that lead to misalignments are not necessarily the ones responsible for morphological transformation in the vast majority of the galaxies in EAGLE.

 Recently, \citet{2020ApJ...894..106K} studied the misalignment channels in Horizon-AGN and found that 35\% of misalignments are merger driven while 65\% are non-merger driven. Although the percentage of misalignments associated with mergers is higher in Horizon-AGN than in EAGLE, the results are qualitatively consistent in that most misalignments are not associated with mergers. The non-merger interactions studied by \citet{2020ApJ...894..106K} include interaction with nearby galaxies (23\%), interaction with dense environments or their central galaxies (21\%), and secular evolution including smooth accretion from neighboring filaments (21\%).

\section{Conclusions}\label{conclu}

We studied the origin and evolution across cosmic time of the kinematic misalignment between star-forming gas and stars of simulated galaxies in the EAGLE simulations. We presented a qualitative comparison with the observational results of \cite{Bryant_2018}, who used galaxies from the SAMI survey. Some interesting points of agreement and disagreement were found. Our results can be summarised as follows.

\begin{itemize}[leftmargin=2.5mm,labelindent=0mm,label={\large\textbullet}]
    \item Overall, EAGLE can reproduce the observational incidence of misalignments in the field and clusters and the dependence on optical colour. 
    \item We found a decrease in the misaligned fraction with increasing stellar mass for galaxies with stellar mass less than $10^{10.8}$ M$_{\odot}$. This trend is not observed in \citet{Bryant_2018}, but in general, the fractions of misaligned galaxies predicted by EAGLE are similar to those obtained with observational data for M$_{*}>10^{10}$M$_{\odot}$. However, a more detailed comparison requires us to make sure the galaxy selection in the simulation resembles that in observations as closely as possible (for example, in the distribution of stellar masses and the gas fraction of galaxies for which a PA measurement of the star-forming gas and stellar components are possible). \cite{2019MNRAS.484..869V} showed that this matching has to be done carefully in order to make meaningful quantitative comparisons between simulations and observations. Hence, we highlight here this requirement which we plan to carry out in the near future. 
    \item We found interesting trends between the compactness of the star-forming gas component of misaligned versus aligned galaxies that are possible to test with observations. Misaligned galaxies display a trend of a smaller half-mass radii ratio than aligned galaxies. This radii ratio has been suggested to trace quenching and has been shown to depend on the environment in observations \citep{2017MNRAS.464..121S}. 
    \item By studying the connection between misalignments and different internal galaxy properties, we found that galaxies that display misalignments are, on average, prolate, dispersion dominated, gas-poor and have small disc contribution to their total kinematic energy. On the contrary, galaxies that remain aligned between two consecutive snapshots, are on average oblate, gas-rich, rotation dominated and have the most significant disc contributions.
    \item Separating between ETGs and LTGs by the $\kappa_{\rm{co}}$, $D/T$ and $v_{\rm{rot}}/\sigma_{0}$ thresholds suggested by \cite{2017MNRAS.472L..45C} and \cite{Thob_2019}, we found that  $\approx 30\%$ of ETGs display misalignments, in agreement with \cite{2011MNRAS.414..968D}.
    \item We demonstrate that external processes, such as galaxy mergers and gas accretion happen in similar frequency in galaxies that remain aligned or become misaligned. This strongly suggests that external processes are not the main source of galaxies displaying misalignments. Instead, the internal galaxy properties that can lead to quick or slow torquing of the star-forming gas are more likely responsible. A good example of this is galaxy mergers: they happen in a similar frequency in aligned and misaligned galaxies at fixed stellar mass. In the case of disky galaxies (which rarely show misalignments), we think the mechanism at play is that the morphology of these galaxies is able to quickly torque the gas and align it with the disk (unless the gas is in a polar or counter-rotating orbit, both of which are stable), while  spheroidal galaxies (which host most of the misalignment events) do this over much longer timescales \citep{2016MNRAS.457..272D}. This is similar to what \citet{voort2015creation} showed for a single elliptical galaxy, but happening in a large population of ellipticals in a cosmological simulation. 
    \item The abrupt and simultaneous decrease in stellar and SF gas mass is more associated with misalignments than mergers. Galaxies that become misaligned seem to preferentially suffer a reduction of their SF gas mass rather than an increase. We studied the tidal field around these galaxies finding that they are more affected by interaction with nearby galaxies, which likely reduces their gas content and leads to misalignments. However, we emphasise that a larger simulation with a comparable resolution to EAGLE is needed to study the evolution of the tidal force parameter over time and its possible relation with misalignments. 
    \item We found that galaxies that display misalignments have larger ex-situ fractions. The latter is a good indicator of the cumulative effect of interactions over the lifespan of galaxies, and hence an indirect evidence of interactions playing a role in driving misalignments.
    
\end{itemize}

Overall, environment appears to play an important role in nurturing the interactions that can lead to the star-forming gas becoming misaligned in the first place. However, the properties internal to galaxies (shape, kinematics, etc.) play a crucial role in determining whether the gas quickly aligns with the stellar component or not. Hence, galaxies that are more triaxial and more dispersion dominated display more misalignments because they are inefficient at realigning the star-forming gas towards the stellar angular momentum vector, i.e., their relaxation time is longer.

\section*{Acknowledgements}

We thank the referee for the insightful comments and suggestions. CC and NP acknowledge support from ANID Basal Project AFB170002 and ANID FONDECYT 1191813. CC is supported by ANID-PFCHA/Doctorado Nacional/2020 - 21202137.  NP gratefully acknowledges support by the ANID BASAL projects ACE210002 and FB210003. CL is funded by the ARC Centre of Excellence for All Sky Astrophysics in 3 Dimensions (ASTRO 3D), through project number CE170100013. CL also thanks the MERAC
Foundation for a Postdoctoral Research Award. 
\newline We acknowledge the Virgo Consortium for making their simulation data available. The EAGLE simulations were performed using the DiRAC-2 facility at Durham, managed by the ICC, and the PRACE facility Curie based in France at TGCC, CEA, Bruyères-le-Ch\^{}atel.
\newline This work was enabled by the following software tools:
\begin{itemize}
    \item \textsc{Python3} \citep{10.5555/1593511}
    \item \textsc{pandas} \citep{mckinney-proc-scipy-2010}
    \item \textsc{NumPy} \citep{harris2020array}
    \item \textsc{SciPy}\citep{2020SciPy-NMeth}
    \item \textsc{Matplotlib} \citep{Hunter:2007}
\end{itemize}
\section*{Data Availability}

The EAGLE database is publicly available at \url{http://icc.dur.ac.uk/Eagle/database.php}. \citet{McAlpine_2016} and \citet{2017arXiv170609899T} provide a detailed guide to access and query it. 
Other data underlying this article are available from the corresponding author on reasonable request.



\bibliographystyle{mnras}
\bibliography{example} 

\begin{thebibliography}{}
\makeatletter
\relax
\def\mn@urlcharsother{\let\do\@makeother \do\$\do\&\do\#\do\^\do\_\do\%\do\~}
\def\mn@doi{\begingroup\mn@urlcharsother \@ifnextchar [ {\mn@doi@}
  {\mn@doi@[]}}
\def\mn@doi@[#1]#2{\def\@tempa{#1}\ifx\@tempa\@empty \href
  {http://dx.doi.org/#2} {doi:#2}\else \href {http://dx.doi.org/#2} {#1}\fi
  \endgroup}
\def\mn@eprint#1#2{\mn@eprint@#1:#2::\@nil}
\def\mn@eprint@arXiv#1{\href {http://arxiv.org/abs/#1} {{\tt arXiv:#1}}}
\def\mn@eprint@dblp#1{\href {http://dblp.uni-trier.de/rec/bibtex/#1.xml}
  {dblp:#1}}
\def\mn@eprint@#1:#2:#3:#4\@nil{\def\@tempa {#1}\def\@tempb {#2}\def\@tempc
  {#3}\ifx \@tempc \@empty \let \@tempc \@tempb \let \@tempb \@tempa \fi \ifx
  \@tempb \@empty \def\@tempb {arXiv}\fi \@ifundefined
  {mn@eprint@\@tempb}{\@tempb:\@tempc}{\expandafter \expandafter \csname
  mn@eprint@\@tempb\endcsname \expandafter{\@tempc}}}

\bibitem[\protect\citeauthoryear{{Argudo-Fern{\'a}ndez}
  et~al.,}{{Argudo-Fern{\'a}ndez} et~al.}{2015}]{2015A&A...578A.110A}
{Argudo-Fern{\'a}ndez} M.,  et~al., 2015, \mn@doi [\aap]
  {10.1051/0004-6361/201526016}, \href
  {https://ui.adsabs.harvard.edu/abs/2015A&A...578A.110A} {578, A110}

\bibitem[\protect\citeauthoryear{{Bauer}, {Drory}  \& {Hill}}{{Bauer}
  et~al.}{2007}]{2007ASSP....3..487B}
{Bauer} A.~E.,  {Drory} N.,   {Hill} G.~J.,  2007, \mn@doi [Astrophysics and
  Space Science Proceedings] {10.1007/978-1-4020-5573-7\_84}, \href
  {https://ui.adsabs.harvard.edu/abs/2007ASSP....3..487B} {3, 487}

\bibitem[\protect\citeauthoryear{{B{\'e}thermin} et~al.,}{{B{\'e}thermin}
  et~al.}{2015}]{2015A&A...573A.113B}
{B{\'e}thermin} M.,  et~al., 2015, \mn@doi [\aap]
  {10.1051/0004-6361/201425031}, \href
  {https://ui.adsabs.harvard.edu/abs/2015A&A...573A.113B} {573, A113}

\bibitem[\protect\citeauthoryear{{Booth} \& {Schaye}}{{Booth} \&
  {Schaye}}{2009}]{2009MNRAS.398...53B}
{Booth} C.~M.,  {Schaye} J.,  2009, \mn@doi [\mnras]
  {10.1111/j.1365-2966.2009.15043.x}, \href
  {https://ui.adsabs.harvard.edu/abs/2009MNRAS.398...53B} {398, 53}

\bibitem[\protect\citeauthoryear{{Bravo}, {Robotham}, {Lagos}, {Davies},
  {Bellstedt}  \& {Thorne}}{{Bravo} et~al.}{2021}]{Bravo2021}
{Bravo} M.,  {Robotham} A. S.~G.,  {Lagos} C. d.~P.,  {Davies} L. J.~M.,
  {Bellstedt} S.,   {Thorne} J.~E.,  2021, arXiv e-prints, \href
  {https://ui.adsabs.harvard.edu/abs/2021arXiv210611829B} {p. arXiv:2106.11829}

\bibitem[\protect\citeauthoryear{{Bryan}, {Kay}, {Duffy}, {Schaye}, {Dalla
  Vecchia}  \& {Booth}}{{Bryan} et~al.}{2013}]{2013MNRAS.429.3316B}
{Bryan} S.~E.,  {Kay} S.~T.,  {Duffy} A.~R.,  {Schaye} J.,  {Dalla Vecchia} C.,
    {Booth} C.~M.,  2013, \mn@doi [\mnras] {10.1093/mnras/sts587}, \href
  {https://ui.adsabs.harvard.edu/abs/2013MNRAS.429.3316B} {429, 3316}

\bibitem[\protect\citeauthoryear{{Bryant} et~al.,}{{Bryant}
  et~al.}{2015}]{2015MNRAS.447.2857B}
{Bryant} J.~J.,  et~al., 2015, \mn@doi [\mnras] {10.1093/mnras/stu2635}, \href
  {https://ui.adsabs.harvard.edu/abs/2015MNRAS.447.2857B} {447, 2857}

\bibitem[\protect\citeauthoryear{Bryant et~al.,}{Bryant
  et~al.}{2018}]{Bryant_2018}
Bryant J.~J.,  et~al., 2018, \mn@doi [\mnras] {10.1093/mnras/sty3122}, 483,
  458–479

\bibitem[\protect\citeauthoryear{{Bundy} et~al.,}{{Bundy}
  et~al.}{2015}]{2015ApJ...798....7B}
{Bundy} K.,  et~al., 2015, \mn@doi [\apj] {10.1088/0004-637X/798/1/7}, \href
  {https://ui.adsabs.harvard.edu/abs/2015ApJ...798....7B} {798, 7}

\bibitem[\protect\citeauthoryear{{Cappellari} et~al.,}{{Cappellari}
  et~al.}{2011}]{2011MNRAS.413..813C}
{Cappellari} M.,  et~al., 2011, \mn@doi [\mnras]
  {10.1111/j.1365-2966.2010.18174.x}, \href
  {https://ui.adsabs.harvard.edu/abs/2011MNRAS.413..813C} {413, 813}

\bibitem[\protect\citeauthoryear{Chen et~al.,}{Chen et~al.}{2016}]{Chen_2016}
Chen Y.-M.,  et~al., 2016, \mn@doi [Nature Communications]
  {10.1038/ncomms13269}, 7

\bibitem[\protect\citeauthoryear{Clauwens, Schaye, Franx  \& Bower}{Clauwens
  et~al.}{2018}]{Clauwens_2018}
Clauwens B.,  Schaye J.,  Franx M.,   Bower R.~G.,  2018, \mn@doi [\mnras]
  {10.1093/mnras/sty1229}, 478, 3994–4009

\bibitem[\protect\citeauthoryear{{Cochrane} \& {Best}}{{Cochrane} \&
  {Best}}{2018}]{Cochrane2018}
{Cochrane} R.~K.,  {Best} P.~N.,  2018, \mn@doi [\mnras]
  {10.1093/mnras/sty1708}, \href
  {https://ui.adsabs.harvard.edu/abs/2018MNRAS.480..864C} {480, 864}

\bibitem[\protect\citeauthoryear{Cole, Lacey, Baugh  \& Frenk}{Cole
  et~al.}{2002}]{Cole_2002}
Cole S.,  Lacey C.~G.,  Baugh C.~M.,   Frenk C.~S.,  2002, \mn@doi [\mnras]
  {10.1046/j.1365-8711.2000.03879.x}, 319, 168–204

\bibitem[\protect\citeauthoryear{{Collacchioni}, {Lagos}, {Mitchell}, {Schaye},
  {Wisnioski}, {Cora}  \& {Correa}}{{Collacchioni}
  et~al.}{2020}]{2020MNRAS.495.2827C}
{Collacchioni} F.,  {Lagos} C. D.~P.,  {Mitchell} P.~D.,  {Schaye} J.,
  {Wisnioski} E.,  {Cora} S.~A.,   {Correa} C.~A.,  2020, \mn@doi [\mnras]
  {10.1093/mnras/staa1334}, \href
  {https://ui.adsabs.harvard.edu/abs/2020MNRAS.495.2827C} {495, 2827}

\bibitem[\protect\citeauthoryear{{Cora}}{{Cora}}{2006}]{2006MNRAS.368.1540C}
{Cora} S.~A.,  2006, \mn@doi [\mnras] {10.1111/j.1365-2966.2006.10271.x}, \href
  {https://ui.adsabs.harvard.edu/abs/2006MNRAS.368.1540C} {368, 1540}

\bibitem[\protect\citeauthoryear{{Correa}, {Schaye}, {Clauwens}, {Bower},
  {Crain}, {Schaller}, {Theuns}  \& {Thob}}{{Correa}
  et~al.}{2017}]{2017MNRAS.472L..45C}
{Correa} C.~A.,  {Schaye} J.,  {Clauwens} B.,  {Bower} R.~G.,  {Crain} R.~A.,
  {Schaller} M.,  {Theuns} T.,   {Thob} A. C.~R.,  2017, \mn@doi [\mnras]
  {10.1093/mnrasl/slx133}, \href
  {https://ui.adsabs.harvard.edu/abs/2017MNRAS.472L..45C} {472, L45}

\bibitem[\protect\citeauthoryear{{Crain} et~al.,}{{Crain}
  et~al.}{2009}]{2009MNRAS.399.1773C}
{Crain} R.~A.,  et~al., 2009, \mn@doi [\mnras]
  {10.1111/j.1365-2966.2009.15402.x}, \href
  {https://ui.adsabs.harvard.edu/abs/2009MNRAS.399.1773C} {399, 1773}

\bibitem[\protect\citeauthoryear{Crain, McCarthy, Frenk, Theuns  \&
  Schaye}{Crain et~al.}{2010}]{Crain_2010}
Crain R.~A.,  McCarthy I.~G.,  Frenk C.~S.,  Theuns T.,   Schaye J.,  2010,
  \mn@doi [\mnras] {10.1111/j.1365-2966.2010.16985.x}, 407, 1403–1422

\bibitem[\protect\citeauthoryear{Crain et~al.,}{Crain
  et~al.}{2015}]{Crain_2015}
Crain R.~A.,  et~al., 2015, \mn@doi [\mnras] {10.1093/mnras/stv725}, 450,
  1937–1961

\bibitem[\protect\citeauthoryear{{Crain} et~al.,}{{Crain}
  et~al.}{2017}]{2017MNRAS.464.4204C}
{Crain} R.~A.,  et~al., 2017, \mn@doi [\mnras] {10.1093/mnras/stw2586}, \href
  {https://ui.adsabs.harvard.edu/abs/2017MNRAS.464.4204C} {464, 4204}

\bibitem[\protect\citeauthoryear{{Croom} et~al.,}{{Croom}
  et~al.}{2012}]{2012MNRAS.421..872C}
{Croom} S.~M.,  et~al., 2012, \mn@doi [\mnras]
  {10.1111/j.1365-2966.2011.20365.x}, \href
  {https://ui.adsabs.harvard.edu/abs/2012MNRAS.421..872C} {421, 872}

\bibitem[\protect\citeauthoryear{{Dalla Vecchia} \& {Schaye}}{{Dalla Vecchia}
  \& {Schaye}}{2012}]{2012MNRAS.426..140D}
{Dalla Vecchia} C.,  {Schaye} J.,  2012, \mn@doi [\mnras]
  {10.1111/j.1365-2966.2012.21704.x}, \href
  {https://ui.adsabs.harvard.edu/abs/2012MNRAS.426..140D} {426, 140}

\bibitem[\protect\citeauthoryear{{Davis} \& {Bureau}}{{Davis} \&
  {Bureau}}{2016}]{2016MNRAS.457..272D}
{Davis} T.~A.,  {Bureau} M.,  2016, \mn@doi [\mnras] {10.1093/mnras/stv2998},
  \href {https://ui.adsabs.harvard.edu/abs/2016MNRAS.457..272D} {457, 272}

\bibitem[\protect\citeauthoryear{{Davis}, {Efstathiou}, {Frenk}  \&
  {White}}{{Davis} et~al.}{1985}]{1985ApJ...292..371D}
{Davis} M.,  {Efstathiou} G.,  {Frenk} C.~S.,   {White} S.~D.~M.,  1985,
  \mn@doi [\apj] {10.1086/163168}, \href
  {https://ui.adsabs.harvard.edu/abs/1985ApJ...292..371D} {292, 371}

\bibitem[\protect\citeauthoryear{{Davis} et~al.,}{{Davis}
  et~al.}{2011}]{2011MNRAS.414..968D}
{Davis} T.~A.,  et~al., 2011, \mn@doi [\mnras]
  {10.1111/j.1365-2966.2011.18284.x}, \href
  {https://ui.adsabs.harvard.edu/abs/2011MNRAS.414..968D} {414, 968}

\bibitem[\protect\citeauthoryear{{Davison}, {Norris}, {Pfeffer}, {Davies}  \&
  {Crain}}{{Davison} et~al.}{2020}]{2020MNRAS.497...81D}
{Davison} T.~A.,  {Norris} M.~A.,  {Pfeffer} J.~L.,  {Davies} J.~J.,   {Crain}
  R.~A.,  2020, \mn@doi [\mnras] {10.1093/mnras/staa1816}, \href
  {https://ui.adsabs.harvard.edu/abs/2020MNRAS.497...81D} {497, 81}

\bibitem[\protect\citeauthoryear{{DeFelippis}, {Genel}, {Bryan}  \&
  {Fall}}{{DeFelippis} et~al.}{2017}]{2017ApJ...841...16D}
{DeFelippis} D.,  {Genel} S.,  {Bryan} G.~L.,   {Fall} S.~M.,  2017, \mn@doi
  [\apj] {10.3847/1538-4357/aa6dfc}, \href
  {https://ui.adsabs.harvard.edu/abs/2017ApJ...841...16D} {841, 16}

\bibitem[\protect\citeauthoryear{{Dolag}, {Borgani}, {Murante}  \&
  {Springel}}{{Dolag} et~al.}{2009}]{2009MNRAS.399..497D}
{Dolag} K.,  {Borgani} S.,  {Murante} G.,   {Springel} V.,  2009, \mn@doi
  [\mnras] {10.1111/j.1365-2966.2009.15034.x}, \href
  {https://ui.adsabs.harvard.edu/abs/2009MNRAS.399..497D} {399, 497}

\bibitem[\protect\citeauthoryear{{Doroshkevich}}{{Doroshkevich}}{1970}]{1970Afz.....6..581D}
{Doroshkevich} A.~G.,  1970, Astrofizika, \href
  {https://ui.adsabs.harvard.edu/abs/1970Afz.....6..581D} {6, 581}

\bibitem[\protect\citeauthoryear{{Driver} et~al.,}{{Driver}
  et~al.}{2018}]{2018MNRAS.475.2891D}
{Driver} S.~P.,  et~al., 2018, \mn@doi [\mnras] {10.1093/mnras/stx2728}, \href
  {https://ui.adsabs.harvard.edu/abs/2018MNRAS.475.2891D} {475, 2891}

\bibitem[\protect\citeauthoryear{Dubois et~al.,}{Dubois
  et~al.}{2014}]{10.1093/mnras/stu1227}
Dubois Y.,  et~al., 2014, \mn@doi [\mnras] {10.1093/mnras/stu1227}, 444, 1453

\bibitem[\protect\citeauthoryear{Duckworth, Tojeiro, Kraljic, Sgró, Wild,
  Weijmans, Lacerna  \& Drory}{Duckworth et~al.}{2018}]{Duckworth_2018}
Duckworth C.,  Tojeiro R.,  Kraljic K.,  Sgró M.~A.,  Wild V.,  Weijmans
  A.-M.,  Lacerna I.,   Drory N.,  2018, \mn@doi [\mnras]
  {10.1093/mnras/sty3101}, 483, 172–188

\bibitem[\protect\citeauthoryear{Duckworth, Tojeiro  \& Kraljic}{Duckworth
  et~al.}{2019}]{Duckworth2020a}
Duckworth C.,  Tojeiro R.,   Kraljic K.,  2019, \mn@doi [Monthly Notices of the
  Royal Astronomical Society] {10.1093/mnras/stz3575}, 492, 1869

\bibitem[\protect\citeauthoryear{{Duckworth}, {Starkenburg}, {Genel}, {Davis},
  {Habouzit}, {Kraljic}  \& {Tojeiro}}{{Duckworth}
  et~al.}{2020}]{Duckworth2020b}
{Duckworth} C.,  {Starkenburg} T.~K.,  {Genel} S.,  {Davis} T.~A.,  {Habouzit}
  M.,  {Kraljic} K.,   {Tojeiro} R.,  2020, \mn@doi [\mnras]
  {10.1093/mnras/staa1494}, \href
  {https://ui.adsabs.harvard.edu/abs/2020MNRAS.495.4542D} {495, 4542}

\bibitem[\protect\citeauthoryear{{Emsellem} et~al.,}{{Emsellem}
  et~al.}{2007}]{2007MNRAS.379..401E}
{Emsellem} E.,  et~al., 2007, \mn@doi [\mnras]
  {10.1111/j.1365-2966.2007.11752.x}, \href
  {https://ui.adsabs.harvard.edu/abs/2007MNRAS.379..401E} {379, 401}

\bibitem[\protect\citeauthoryear{{Fall} \& {Efstathiou}}{{Fall} \&
  {Efstathiou}}{1980}]{1980MNRAS.193..189F}
{Fall} S.~M.,  {Efstathiou} G.,  1980, \mn@doi [\mnras]
  {10.1093/mnras/193.2.189}, \href
  {https://ui.adsabs.harvard.edu/abs/1980MNRAS.193..189F} {193, 189}

\bibitem[\protect\citeauthoryear{{Fogarty} et~al.,}{{Fogarty}
  et~al.}{2014}]{2014MNRAS.443..485F}
{Fogarty} L.~M.~R.,  et~al., 2014, \mn@doi [\mnras] {10.1093/mnras/stu1165},
  \href {https://ui.adsabs.harvard.edu/abs/2014MNRAS.443..485F} {443, 485}

\bibitem[\protect\citeauthoryear{{Foster} et~al.,}{{Foster}
  et~al.}{2021}]{Foster2021}
{Foster} C.,  et~al., 2021, \mn@doi [\pasa] {10.1017/pasa.2021.25}, \href
  {https://ui.adsabs.harvard.edu/abs/2021PASA...38...31F} {38, e031}

\bibitem[\protect\citeauthoryear{{Furlong} et~al.,}{{Furlong}
  et~al.}{2015}]{2015MNRAS.450.4486F}
{Furlong} M.,  et~al., 2015, \mn@doi [\mnras] {10.1093/mnras/stv852}, \href
  {https://ui.adsabs.harvard.edu/abs/2015MNRAS.450.4486F} {450, 4486}

\bibitem[\protect\citeauthoryear{Garrison-Kimmel et~al.,}{Garrison-Kimmel
  et~al.}{2018}]{10.1093/mnras/sty2513}
Garrison-Kimmel S.,  et~al., 2018, \mn@doi [\mnras] {10.1093/mnras/sty2513},
  481, 4133

\bibitem[\protect\citeauthoryear{{Haardt} \& {Madau}}{{Haardt} \&
  {Madau}}{2001}]{2001cghr.confE..64H}
{Haardt} F.,  {Madau} P.,  2001, in {Neumann} D.~M.,  {Tran} J.~T.~V.,  eds,
  Clusters of Galaxies and the High Redshift Universe Observed in X-rays. p.~64
  (\mn@eprint {arXiv} {astro-ph/0106018})

\bibitem[\protect\citeauthoryear{Harris et~al.,}{Harris
  et~al.}{2020}]{harris2020array}
Harris C.~R.,  et~al., 2020, \mn@doi [Nature] {10.1038/s41586-020-2649-2}, 585,
  357

\bibitem[\protect\citeauthoryear{Hunter}{Hunter}{2007}]{Hunter:2007}
Hunter J.~D.,  2007, \mn@doi [Computing in Science \& Engineering]
  {10.1109/MCSE.2007.55}, 9, 90

\bibitem[\protect\citeauthoryear{{Jenkins} \& {Booth}}{{Jenkins} \&
  {Booth}}{2013}]{2013arXiv1306.5771J}
{Jenkins} A.,  {Booth} S.,  2013, arXiv e-prints, \href
  {https://ui.adsabs.harvard.edu/abs/2013arXiv1306.5771J} {p. arXiv:1306.5771}

\bibitem[\protect\citeauthoryear{Jiang, Helly, Cole  \& Frenk}{Jiang
  et~al.}{2014}]{Jiang_2014}
Jiang L.,  Helly J.~C.,  Cole S.,   Frenk C.~S.,  2014, \mn@doi [\mnras]
  {10.1093/mnras/stu390}, 440, 2115–2135

\bibitem[\protect\citeauthoryear{Jin et~al.,}{Jin et~al.}{2016}]{Jin_2016}
Jin Y.,  et~al., 2016, \mn@doi [\mnras] {10.1093/mnras/stw2055}, 463, 913–926

\bibitem[\protect\citeauthoryear{{Khim}, {Yi}, {Pichon}, {Dubois}, {Devriendt},
  {Choi}, {Bryant}  \& {Croom}}{{Khim} et~al.}{2020a}]{2020arXiv201204659K}
{Khim} D.~J.,  {Yi} S.~K.,  {Pichon} C.,  {Dubois} Y.,  {Devriendt} J.,  {Choi}
  H.,  {Bryant} J.~J.,   {Croom} S.~M.,  2020a, arXiv e-prints, \href
  {https://ui.adsabs.harvard.edu/abs/2020arXiv201204659K} {p. arXiv:2012.04659}

\bibitem[\protect\citeauthoryear{{Khim} et~al.,}{{Khim}
  et~al.}{2020b}]{2020ApJ...894..106K}
{Khim} D.~J.,  et~al., 2020b, \mn@doi [\apj] {10.3847/1538-4357/ab88a9}, \href
  {https://ui.adsabs.harvard.edu/abs/2020ApJ...894..106K} {894, 106}

\bibitem[\protect\citeauthoryear{{Krajnovi{\'c}}, {Cappellari}, {de Zeeuw}  \&
  {Copin}}{{Krajnovi{\'c}} et~al.}{2006}]{2006MNRAS.366..787K}
{Krajnovi{\'c}} D.,  {Cappellari} M.,  {de Zeeuw} P.~T.,   {Copin} Y.,  2006,
  \mn@doi [\mnras] {10.1111/j.1365-2966.2005.09902.x}, \href
  {https://ui.adsabs.harvard.edu/abs/2006MNRAS.366..787K} {366, 787}

\bibitem[\protect\citeauthoryear{Lagos, Cora  \& Padilla}{Lagos
  et~al.}{2008}]{10.1111/j.1365-2966.2008.13456.x}
Lagos C. d.~P.,  Cora S.~A.,   Padilla N.~D.,  2008, \mn@doi [\mnras]
  {10.1111/j.1365-2966.2008.13456.x}, 388, 587

\bibitem[\protect\citeauthoryear{{Lagos}, {Padilla}, {Davis}, {Lacey}, {Baugh},
  {Gonzalez-Perez}, {Zwaan}  \& {Contreras}}{{Lagos}
  et~al.}{2015}]{2015MNRAS.448.1271L}
{Lagos} C. d.~P.,  {Padilla} N.~D.,  {Davis} T.~A.,  {Lacey} C.~G.,  {Baugh}
  C.~M.,  {Gonzalez-Perez} V.,  {Zwaan} M.~A.,   {Contreras} S.,  2015, \mn@doi
  [\mnras] {10.1093/mnras/stu2763}, \href
  {https://ui.adsabs.harvard.edu/abs/2015MNRAS.448.1271L} {448, 1271}

\bibitem[\protect\citeauthoryear{{Lagos}, {Theuns}, {Stevens}, {Cortese},
  {Padilla}, {Davis}, {Contreras}  \& {Croton}}{{Lagos}
  et~al.}{2017}]{2017MNRAS.464.3850L}
{Lagos} C. d.~P.,  {Theuns} T.,  {Stevens} A. R.~H.,  {Cortese} L.,  {Padilla}
  N.~D.,  {Davis} T.~A.,  {Contreras} S.,   {Croton} D.,  2017, \mn@doi
  [\mnras] {10.1093/mnras/stw2610}, \href
  {https://ui.adsabs.harvard.edu/abs/2017MNRAS.464.3850L} {464, 3850}

\bibitem[\protect\citeauthoryear{{Lagos}, {Schaye}, {Bah{\'e}}, {Van de Sande},
  {Kay}, {Barnes}, {Davis}  \& {Dalla Vecchia}}{{Lagos}
  et~al.}{2018a}]{2018MNRAS.476.4327L}
{Lagos} C. d.~P.,  {Schaye} J.,  {Bah{\'e}} Y.,  {Van de Sande} J.,  {Kay}
  S.~T.,  {Barnes} D.,  {Davis} T.~A.,   {Dalla Vecchia} C.,  2018a, \mn@doi
  [\mnras] {10.1093/mnras/sty489}, \href
  {https://ui.adsabs.harvard.edu/abs/2018MNRAS.476.4327L} {476, 4327}

\bibitem[\protect\citeauthoryear{Lagos, Schaye, Bahé, Van~de Sande, Kay,
  Barnes, Davis  \& Dalla~Vecchia}{Lagos et~al.}{2018b}]{10.1093/mnras/sty489}
Lagos C. d.~P.,  Schaye J.,  Bahé Y.,  Van~de Sande J.,  Kay S.~T.,  Barnes
  D.,  Davis T.~A.,   Dalla~Vecchia C.,  2018b, \mn@doi [\mnras]
  {10.1093/mnras/sty489}, 476, 4327

\bibitem[\protect\citeauthoryear{{Leja}, {Tacchella}  \& {Conroy}}{{Leja}
  et~al.}{2019}]{Leja2019}
{Leja} J.,  {Tacchella} S.,   {Conroy} C.,  2019, \mn@doi [\apjl]
  {10.3847/2041-8213/ab2f8c}, \href
  {https://ui.adsabs.harvard.edu/abs/2019ApJ...880L...9L} {880, L9}

\bibitem[\protect\citeauthoryear{McAlpine et~al.,}{McAlpine
  et~al.}{2016}]{McAlpine_2016}
McAlpine S.,  et~al., 2016, \mn@doi [Astronomy and Computing]
  {10.1016/j.ascom.2016.02.004}, 15, 72–89

\bibitem[\protect\citeauthoryear{McKinney}{McKinney}{2010}]{mckinney-proc-scipy-2010}
McKinney W.,  2010, in van~der Walt S.,  Millman J.,  eds, Proceedings of the
  9th Python in Science Conference. pp 51 -- 56

\bibitem[\protect\citeauthoryear{{Mitchell}, {Lacey}, {Cole}  \&
  {Baugh}}{{Mitchell} et~al.}{2014}]{2014MNRAS.444.2637M}
{Mitchell} P.~D.,  {Lacey} C.~G.,  {Cole} S.,   {Baugh} C.~M.,  2014, \mn@doi
  [\mnras] {10.1093/mnras/stu1639}, \href
  {https://ui.adsabs.harvard.edu/abs/2014MNRAS.444.2637M} {444, 2637}

\bibitem[\protect\citeauthoryear{{Mo}, {Mao}  \& {White}}{{Mo}
  et~al.}{1998}]{1998MNRAS.295..319M}
{Mo} H.~J.,  {Mao} S.,   {White} S. D.~M.,  1998, \mn@doi [\mnras]
  {10.1046/j.1365-8711.1998.01227.x}, \href
  {https://ui.adsabs.harvard.edu/abs/1998MNRAS.295..319M} {295, 319}

\bibitem[\protect\citeauthoryear{Nelson et~al.,}{Nelson
  et~al.}{2015}]{Nelson_2015}
Nelson D.,  et~al., 2015, \mn@doi [Astronomy and Computing]
  {10.1016/j.ascom.2015.09.003}, 13, 12–37

\bibitem[\protect\citeauthoryear{Nelson et~al.,}{Nelson
  et~al.}{2019}]{Nelson2019TheIS}
Nelson D.,  et~al., 2019, Computational Astrophysics and Cosmology, 6, 1

\bibitem[\protect\citeauthoryear{{Padilla}, {Salazar-Albornoz}, {Contreras},
  {Cora}  \& {Ruiz}}{{Padilla} et~al.}{2014}]{2014MNRAS.443.2801P}
{Padilla} N.~D.,  {Salazar-Albornoz} S.,  {Contreras} S.,  {Cora} S.~A.,
  {Ruiz} A.~N.,  2014, \mn@doi [\mnras] {10.1093/mnras/stu1321}, \href
  {https://ui.adsabs.harvard.edu/abs/2014MNRAS.443.2801P} {443, 2801}

\bibitem[\protect\citeauthoryear{{Pedrosa} \& {Tissera}}{{Pedrosa} \&
  {Tissera}}{2015}]{2015A&A...584A..43P}
{Pedrosa} S.~E.,  {Tissera} P.~B.,  2015, \mn@doi [\aap]
  {10.1051/0004-6361/201526440}, \href
  {https://ui.adsabs.harvard.edu/abs/2015A&A...584A..43P} {584, A43}

\bibitem[\protect\citeauthoryear{{Peebles}}{{Peebles}}{1969}]{1969ApJ...155..393P}
{Peebles} P.~J.~E.,  1969, \mn@doi [\apj] {10.1086/149876}, \href
  {https://ui.adsabs.harvard.edu/abs/1969ApJ...155..393P} {155, 393}

\bibitem[\protect\citeauthoryear{{Peng} et~al.,}{{Peng}
  et~al.}{2010}]{Peng2010}
{Peng} Y.-j.,  et~al., 2010, \mn@doi [\apj] {10.1088/0004-637X/721/1/193},
  \href {https://ui.adsabs.harvard.edu/abs/2010ApJ...721..193P} {721, 193}

\bibitem[\protect\citeauthoryear{{Planck Collaboration} et~al.,}{{Planck
  Collaboration} et~al.}{2015}]{refId0}
{Planck Collaboration} et~al., 2015, \mn@doi [A\&A]
  {10.1051/0004-6361/201525830}, 594, A13

\bibitem[\protect\citeauthoryear{{Qu} et~al.,}{{Qu}
  et~al.}{2017}]{2017MNRAS.464.1659Q}
{Qu} Y.,  et~al., 2017, \mn@doi [\mnras] {10.1093/mnras/stw2437}, \href
  {https://ui.adsabs.harvard.edu/abs/2017MNRAS.464.1659Q} {464, 1659}

\bibitem[\protect\citeauthoryear{{Rosas-Guevara} et~al.,}{{Rosas-Guevara}
  et~al.}{2015}]{2015MNRAS.454.1038R}
{Rosas-Guevara} Y.~M.,  et~al., 2015, \mn@doi [\mnras] {10.1093/mnras/stv2056},
  \href {https://ui.adsabs.harvard.edu/abs/2015MNRAS.454.1038R} {454, 1038}

\bibitem[\protect\citeauthoryear{{Sales}, {Navarro}, {Theuns}, {Schaye},
  {White}, {Frenk}, {Crain}  \& {Dalla Vecchia}}{{Sales}
  et~al.}{2012}]{2012MNRAS.423.1544S}
{Sales} L.~V.,  {Navarro} J.~F.,  {Theuns} T.,  {Schaye} J.,  {White} S. D.~M.,
   {Frenk} C.~S.,  {Crain} R.~A.,   {Dalla Vecchia} C.,  2012, \mn@doi [\mnras]
  {10.1111/j.1365-2966.2012.20975.x}, \href
  {https://ui.adsabs.harvard.edu/abs/2012MNRAS.423.1544S} {423, 1544}

\bibitem[\protect\citeauthoryear{{S{\'a}nchez} et~al.,}{{S{\'a}nchez}
  et~al.}{2012}]{2012A&A...538A...8S}
{S{\'a}nchez} S.~F.,  et~al., 2012, \mn@doi [\aap]
  {10.1051/0004-6361/201117353}, \href
  {https://ui.adsabs.harvard.edu/abs/2012A&A...538A...8S} {538, A8}

\bibitem[\protect\citeauthoryear{{Schaefer} et~al.,}{{Schaefer}
  et~al.}{2017}]{2017MNRAS.464..121S}
{Schaefer} A.~L.,  et~al., 2017, \mn@doi [\mnras] {10.1093/mnras/stw2289},
  \href {https://ui.adsabs.harvard.edu/abs/2017MNRAS.464..121S} {464, 121}

\bibitem[\protect\citeauthoryear{Schaller, Dalla~Vecchia, Schaye, Bower,
  Theuns, Crain, Furlong  \& McCarthy}{Schaller
  et~al.}{2015}]{10.1093/mnras/stv2169}
Schaller M.,  Dalla~Vecchia C.,  Schaye J.,  Bower R.~G.,  Theuns T.,  Crain
  R.~A.,  Furlong M.,   McCarthy I.~G.,  2015, \mn@doi [\mnras]
  {10.1093/mnras/stv2169}, 454, 2277

\bibitem[\protect\citeauthoryear{{Schaye} \& {Dalla Vecchia}}{{Schaye} \&
  {Dalla Vecchia}}{2008}]{2008MNRAS.383.1210S}
{Schaye} J.,  {Dalla Vecchia} C.,  2008, \mn@doi [\mnras]
  {10.1111/j.1365-2966.2007.12639.x}, \href
  {https://ui.adsabs.harvard.edu/abs/2008MNRAS.383.1210S} {383, 1210}

\bibitem[\protect\citeauthoryear{{Schaye} et~al.,}{{Schaye}
  et~al.}{2015}]{2015MNRAS.446..521S}
{Schaye} J.,  et~al., 2015, \mn@doi [\mnras] {10.1093/mnras/stu2058}, \href
  {https://ui.adsabs.harvard.edu/abs/2015MNRAS.446..521S} {446, 521}

\bibitem[\protect\citeauthoryear{{Serra} et~al.,}{{Serra}
  et~al.}{2012}]{2012MNRAS.422.1835S}
{Serra} P.,  et~al., 2012, \mn@doi [\mnras] {10.1111/j.1365-2966.2012.20219.x},
  \href {https://ui.adsabs.harvard.edu/abs/2012MNRAS.422.1835S} {422, 1835}

\bibitem[\protect\citeauthoryear{{Springel}}{{Springel}}{2005}]{2005MNRAS.364.1105S}
{Springel} V.,  2005, \mn@doi [\mnras] {10.1111/j.1365-2966.2005.09655.x},
  \href {https://ui.adsabs.harvard.edu/abs/2005MNRAS.364.1105S} {364, 1105}

\bibitem[\protect\citeauthoryear{{Springel}, {White}, {Tormen}  \&
  {Kauffmann}}{{Springel} et~al.}{2001}]{2001MNRAS.328..726S}
{Springel} V.,  {White} S. D.~M.,  {Tormen} G.,   {Kauffmann} G.,  2001,
  \mn@doi [\mnras] {10.1046/j.1365-8711.2001.04912.x}, \href
  {https://ui.adsabs.harvard.edu/abs/2001MNRAS.328..726S} {328, 726}

\bibitem[\protect\citeauthoryear{Starkenburg, Sales, Genel, Manzano-King,
  Canalizo  \& Hernquist}{Starkenburg et~al.}{2019}]{Starkenburg_2019}
Starkenburg T.~K.,  Sales L.~V.,  Genel S.,  Manzano-King C.,  Canalizo G.,
  Hernquist L.,  2019, \mn@doi [\apj] {10.3847/1538-4357/ab2128}, 878, 143

\bibitem[\protect\citeauthoryear{{Tecce}, {Cora}, {Tissera}, {Abadi}  \&
  {Lagos}}{{Tecce} et~al.}{2010}]{2010MNRAS.408.2008T}
{Tecce} T.~E.,  {Cora} S.~A.,  {Tissera} P.~B.,  {Abadi} M.~G.,   {Lagos} C.
  D.~P.,  2010, \mn@doi [\mnras] {10.1111/j.1365-2966.2010.17262.x}, \href
  {https://ui.adsabs.harvard.edu/abs/2010MNRAS.408.2008T} {408, 2008}

\bibitem[\protect\citeauthoryear{{Teklu}, {Remus}, {Dolag}, {Beck}, {Burkert},
  {Schmidt}, {Schulze}  \& {Steinborn}}{{Teklu}
  et~al.}{2015}]{2015ApJ...812...29T}
{Teklu} A.~F.,  {Remus} R.-S.,  {Dolag} K.,  {Beck} A.~M.,  {Burkert} A.,
  {Schmidt} A.~S.,  {Schulze} F.,   {Steinborn} L.~K.,  2015, \mn@doi [\apj]
  {10.1088/0004-637X/812/1/29}, \href
  {https://ui.adsabs.harvard.edu/abs/2015ApJ...812...29T} {812, 29}

\bibitem[\protect\citeauthoryear{{The EAGLE team}}{{The EAGLE
  team}}{2017}]{2017arXiv170609899T}
{The EAGLE team} 2017, arXiv e-prints, \href
  {https://ui.adsabs.harvard.edu/abs/2017arXiv170609899T} {p. arXiv:1706.09899}

\bibitem[\protect\citeauthoryear{Thob et~al.,}{Thob et~al.}{2019}]{Thob_2019}
Thob A. C.~R.,  et~al., 2019, \mn@doi [\mnras] {10.1093/mnras/stz448}, 485,
  972–987

\bibitem[\protect\citeauthoryear{{Trayford} \& {Schaye}}{{Trayford} \&
  {Schaye}}{2019}]{2019MNRAS.485.5715T}
{Trayford} J.~W.,  {Schaye} J.,  2019, \mn@doi [\mnras] {10.1093/mnras/stz757},
  \href {https://ui.adsabs.harvard.edu/abs/2019MNRAS.485.5715T} {485, 5715}

\bibitem[\protect\citeauthoryear{{Trayford}, {Theuns}, {Bower}, {Crain},
  {Lagos}, {Schaller}  \& {Schaye}}{{Trayford} et~al.}{2016}]{Trayford2016}
{Trayford} J.~W.,  {Theuns} T.,  {Bower} R.~G.,  {Crain} R.~A.,  {Lagos} C.
  d.~P.,  {Schaller} M.,   {Schaye} J.,  2016, \mn@doi [\mnras]
  {10.1093/mnras/stw1230}, \href
  {https://ui.adsabs.harvard.edu/abs/2016MNRAS.460.3925T} {460, 3925}

\bibitem[\protect\citeauthoryear{{Trayford}, {Frenk}, {Theuns}, {Schaye}  \&
  {Correa}}{{Trayford} et~al.}{2019}]{2019MNRAS.483..744T}
{Trayford} J.~W.,  {Frenk} C.~S.,  {Theuns} T.,  {Schaye} J.,   {Correa} C.,
  2019, \mn@doi [\mnras] {10.1093/mnras/sty2860}, \href
  {https://ui.adsabs.harvard.edu/abs/2019MNRAS.483..744T} {483, 744}

\bibitem[\protect\citeauthoryear{Van~Rossum \& Drake}{Van~Rossum \&
  Drake}{2009}]{10.5555/1593511}
Van~Rossum G.,  Drake F.~L.,  2009, Python 3 Reference Manual.
CreateSpace, Scotts Valley, CA

\bibitem[\protect\citeauthoryear{Virtanen et~al.,}{Virtanen
  et~al.}{2020}]{2020SciPy-NMeth}
Virtanen P.,  et~al., 2020, \mn@doi [Nature Methods]
  {10.1038/s41592-019-0686-2}, \href {https://rdcu.be/b08Wh} {17, 261}

\bibitem[\protect\citeauthoryear{{Walo-Mart{\'\i}n}, {Falc{\'o}n-Barroso},
  {Dalla Vecchia}, {P{\'e}rez}  \& {Negri}}{{Walo-Mart{\'\i}n}
  et~al.}{2020}]{2020MNRAS.494.5652W}
{Walo-Mart{\'\i}n} D.,  {Falc{\'o}n-Barroso} J.,  {Dalla Vecchia} C.,
  {P{\'e}rez} I.,   {Negri} A.,  2020, \mn@doi [\mnras]
  {10.1093/mnras/staa1066}, \href
  {https://ui.adsabs.harvard.edu/abs/2020MNRAS.494.5652W} {494, 5652}

\bibitem[\protect\citeauthoryear{{White}}{{White}}{1984}]{1984ApJ...286...38W}
{White} S.~D.~M.,  1984, \mn@doi [\apj] {10.1086/162573}, \href
  {https://ui.adsabs.harvard.edu/abs/1984ApJ...286...38W} {286, 38}

\bibitem[\protect\citeauthoryear{{White} \& {Rees}}{{White} \&
  {Rees}}{1978}]{1978MNRAS.183..341W}
{White} S.~D.~M.,  {Rees} M.~J.,  1978, \mn@doi [\mnras]
  {10.1093/mnras/183.3.341}, \href
  {https://ui.adsabs.harvard.edu/abs/1978MNRAS.183..341W} {183, 341}

\bibitem[\protect\citeauthoryear{{Wiersma}, {Schaye}  \& {Smith}}{{Wiersma}
  et~al.}{2009a}]{2009MNRAS.393...99W}
{Wiersma} R. P.~C.,  {Schaye} J.,   {Smith} B.~D.,  2009a, \mn@doi [\mnras]
  {10.1111/j.1365-2966.2008.14191.x}, \href
  {https://ui.adsabs.harvard.edu/abs/2009MNRAS.393...99W} {393, 99}

\bibitem[\protect\citeauthoryear{Wiersma, Schaye, Theuns, Dalla~Vecchia  \&
  Tornatore}{Wiersma et~al.}{2009b}]{Wiersma_2009b}
Wiersma R. P.~C.,  Schaye J.,  Theuns T.,  Dalla~Vecchia C.,   Tornatore L.,
  2009b, \mn@doi [\mnras] {10.1111/j.1365-2966.2009.15331.x}, 399, 574–600

\bibitem[\protect\citeauthoryear{{Wright}, {Lagos}, {Davies}, {Power},
  {Trayford}  \& {Wong}}{{Wright} et~al.}{2019}]{2019MNRAS.487.3740W}
{Wright} R.~J.,  {Lagos} C. d.~P.,  {Davies} L. J.~M.,  {Power} C.,  {Trayford}
  J.~W.,   {Wong} O.~I.,  2019, \mn@doi [\mnras] {10.1093/mnras/stz1410}, \href
  {https://ui.adsabs.harvard.edu/abs/2019MNRAS.487.3740W} {487, 3740}

\bibitem[\protect\citeauthoryear{Zavala et~al.,}{Zavala
  et~al.}{2016}]{10.1093/mnras/stw1286}
Zavala J.,  et~al., 2016, \mn@doi [\mnras] {10.1093/mnras/stw1286}, 460, 4466

\bibitem[\protect\citeauthoryear{Zjupa \& Springel}{Zjupa \&
  Springel}{2016}]{10.1093/mnras/stw2945}
Zjupa J.,  Springel V.,  2016, \mn@doi [\mnras] {10.1093/mnras/stw2945}, 466,
  1625

\bibitem[\protect\citeauthoryear{{de Zeeuw} et~al.,}{{de Zeeuw}
  et~al.}{2002}]{2002MNRAS.329..513D}
{de Zeeuw} P.~T.,  et~al., 2002, \mn@doi [\mnras]
  {10.1046/j.1365-8711.2002.05059.x}, \href
  {https://ui.adsabs.harvard.edu/abs/2002MNRAS.329..513D} {329, 513}

\bibitem[\protect\citeauthoryear{van~de Sande et~al.,}{van~de Sande
  et~al.}{2017}]{10.1093/mnras/stx1751}
van~de Sande J.,  et~al., 2017, \mn@doi [\mnras] {10.1093/mnras/stx1751}, 472,
  1272

\bibitem[\protect\citeauthoryear{{van de Sande} et~al.,}{{van de Sande}
  et~al.}{2019}]{2019MNRAS.484..869V}
{van de Sande} J.,  et~al., 2019, \mn@doi [\mnras] {10.1093/mnras/sty3506},
  \href {https://ui.adsabs.harvard.edu/abs/2019MNRAS.484..869V} {484, 869}

\bibitem[\protect\citeauthoryear{van~de Voort, Davis, Keres, Quataert,
  Faucher-Giguere  \& Hopkins}{van~de Voort et~al.}{2015}]{voort2015creation}
van~de Voort F.,  Davis T.~A.,  Keres D.,  Quataert E.,  Faucher-Giguere C.-A.,
    Hopkins P.~F.,  2015, The creation and persistence of a misaligned gas disc
  in a simulated early-type galaxy (\mn@eprint {arXiv} {1504.03685})

\bibitem[\protect\citeauthoryear{{van den Bosch}, {Abel}, {Croft}, {Hernquist}
  \& {White}}{{van den Bosch} et~al.}{2002}]{2002ApJ...576...21V}
{van den Bosch} F.~C.,  {Abel} T.,  {Croft} R. A.~C.,  {Hernquist} L.,
  {White} S. D.~M.,  2002, \mn@doi [\apj] {10.1086/341619}, \href
  {https://ui.adsabs.harvard.edu/abs/2002ApJ...576...21V} {576, 21}

\makeatother
\end{thebibliography}




\appendix
\section{}

\begin{table}
\centering
\begin{adjustbox}{width=8.49cm}
\begin{tabular}{cccc}

\cline{2-4}
                        & \multicolumn{3}{c}{{Percentage of galaxies below M$_{*}$ = $10^{9.5}$ M$_{\odot}$ (\%)}} \\ \hline
\multicolumn{1}{c}{{Population label}} & ${z=0.87-1.00
}$ & ${z=0.50-0.62}$ & ${z=0.00-0.10}$  \\ \hline \hline

\multicolumn{1}{l}{Remained aligned }  &  27 $\pm$ 0.6     &  26 $\pm$ 0.5  & 30 $\pm$ 0.6  \\ \hline
\multicolumn{1}{l}{Remained misaligned} & 72 $\pm$ 2.3      &  63 $\pm$ 2.7     &    52 $\pm$ 2.9   \\ \hline
\multicolumn{1}{l}{Aligned} &  57 $\pm$ 1.5     &  57 $\pm$ 2.0     &    55 $\pm$ 2.6    \\ \hline

\multicolumn{1}{l}{Misaligned} &  61 $\pm$ 2.0      &   56 $\pm$ 2.0    &   53 $\pm$ 2.4    \\ \hline

\end{tabular}
\end{adjustbox}
  \caption{Percentage of low-mass galaxies (M$_{*} < 10^{9.5}$ M$_{\odot}$) within each sample.}
  \label{anex1}
\end{table}

\begin{table}
\centering
\begin{adjustbox}{width=8.49cm}
\begin{tabular}{cccc}

\cline{2-4}
                        & \multicolumn{3}{c}{{Percentage of galaxies below $f_{\rm{gas}}$ = 0.1 (\%)}} \\ \hline 
\multicolumn{1}{c}{{Population label}} & ${z=0.87-1.00
}$ & ${z=0.50-0.62}$ & ${z=0.00-0.10}$  \\ \hline \hline

\multicolumn{1}{l}{Remained aligned }  &  3 $\pm$ 0.2     &  8 $\pm$ 0.3  & 26 $\pm$ 0.5  \\ \hline
\multicolumn{1}{l}{Remained misaligned} & 3 $\pm$ 0.4      &  20 $\pm$ 1.3     &    63 $\pm$ 3.2   \\ \hline
\multicolumn{1}{l}{Aligned} &  1 $\pm$ 0.2     &  8 $\pm$ 0.6     &    34 $\pm$ 1.9    \\ \hline

\multicolumn{1}{l}{Misaligned} &  5 $\pm$ 0.5      &   16 $\pm$ 0.9    &   52 $\pm$ 2.3    \\ \hline

\end{tabular}
\end{adjustbox}
  \caption{Percentage of galaxies with gas fraction below 0.1 within each sample.}
  \label{anex2}
\end{table}

\begin{table}
\centering
\begin{adjustbox}{width=8.49cm}
\begin{tabular}{cccc}

\cline{2-4}
                        & \multicolumn{3}{c}{{Percentage of galaxies below $T$ = 1/3 (\%)}} \\ \hline
\multicolumn{1}{c}{{Population label}} & ${z=0.87-1.00
}$ & ${z=0.50-0.62}$ & ${z=0.00-0.10}$  \\ \hline \hline

\multicolumn{1}{l}{Remained aligned }  &  61 $\pm$ 0.9     &  70 $\pm$ 1.0  & 78 $\pm$ 1.1  \\ \hline
\multicolumn{1}{l}{Remained misaligned} & 22 $\pm$ 1.1      &  31 $\pm$ 1.7     &    48 $\pm$ 2.7   \\ \hline
\multicolumn{1}{l}{Aligned} &  37 $\pm$ 1.2     &  43 $\pm$ 1.7     &    49 $\pm$ 2.4    \\ \hline

\multicolumn{1}{l}{Misaligned} &  30 $\pm$ 1.2      &   40 $\pm$ 1.6    &   48 $\pm$ 2.2    \\ \hline

\end{tabular}
\end{adjustbox}
  \caption{Percentage of galaxies with stellar triaxiality below 1/3 within each sample.}
  \label{anex3}
\end{table}

\newpage
\begin{table}
\centering
\begin{adjustbox}{width=8.49cm}
\begin{tabular}{cccc}

\cline{2-4}
                        & \multicolumn{3}{c}{{Percentage of galaxies below $\delta$ = 0}} \\ \hline
\multicolumn{1}{c}{{Population label}} & ${z=0.87-1.00
}$ & ${z=0.50-0.62}$ & ${z=0.00-0.10}$  \\ \hline \hline

\multicolumn{1}{l}{Remained aligned }  &  7 $\pm$ 0.3     &  7 $\pm$ 0.2  & 5 $\pm$ 0.2  \\ \hline
\multicolumn{1}{l}{Remained misaligned} & 17 $\pm$ 0.9      &  15 $\pm$ 1.1     &    12 $\pm$ 1.1   \\ \hline
\multicolumn{1}{l}{Aligned} &  13 $\pm$ 0.6     &  13 $\pm$ 0.8     &    13 $\pm$ 1.1    \\ \hline

\multicolumn{1}{l}{Misaligned} &  14 $\pm$ 0.8      &   13 $\pm$ 0.9    &   11 $\pm$ 0.9    \\ \hline

\end{tabular}
\end{adjustbox}
  \caption{Percentage of galaxies with stellar velocity anisotropy below 0 within each sample.}
  \label{anexani}
\end{table}

\begin{table}
\centering
\begin{adjustbox}{width=8.49cm}
\begin{tabular}{cccc}

\cline{2-4}
                        & \multicolumn{3}{c}{{Percentage of galaxies below $D/T$ = 0.45 (\%)}} \\ \hline
\multicolumn{1}{c}{{Population label}} & ${z=0.87-1.00
}$ & ${z=0.50-0.62}$ & ${z=0.00-0.10}$  \\ \hline \hline

\multicolumn{1}{l}{Remained aligned }  &  40 $\pm$ 0.7     &  40 $\pm$ 0.7  & 45 $\pm$ 0.7  \\ \hline
\multicolumn{1}{l}{Remained misaligned} & 92 $\pm$ 2.7      &  91 $\pm$ 3.5     &    84 $\pm$ 4.0   \\ \hline
\multicolumn{1}{l}{Aligned} &  78 $\pm$ 1.9     &  81 $\pm$ 2.6     &    85 $\pm$ 3.5    \\ \hline

\multicolumn{1}{l}{Misaligned} &  81 $\pm$ 2.4      &   76 $\pm$ 2.5    &   78 $\pm$ 3.1    \\ \hline

\end{tabular}
\end{adjustbox}
  \caption{Percentage of galaxies with disc-to-total stellar mass ratio below 0.45 within each sample.}
  \label{anex33}
\end{table}

\begin{table}
\centering
\begin{adjustbox}{width=8.49cm}
\begin{tabular}{cccc}

\cline{2-4}
                        & \multicolumn{3}{c}{{Percentage of galaxies below $v_{\rm{rot}}/\sigma_{0}$ = 0.7 (\%)}} \\ \hline
\multicolumn{1}{c}{{Population label}} & ${z=0.87-1.00
}$ & ${z=0.50-0.62}$ & ${z=0.00-0.10}$  \\ \hline \hline

\multicolumn{1}{l}{Remained aligned }  &  45 $\pm$ 0.8     &  43 $\pm$ 0.7  & 45 $\pm$ 0.7  \\ \hline
\multicolumn{1}{l}{Remained misaligned} & 94 $\pm$ 2.8      &  92 $\pm$ 3.6     &    84 $\pm$ 4.0   \\ \hline
\multicolumn{1}{l}{Aligned} &  83 $\pm$ 2.0     &  84 $\pm$ 2.7     &    86 $\pm$ 3.6    \\ \hline

\multicolumn{1}{l}{Misaligned} &  84 $\pm$ 2.5      &   80 $\pm$ 2.6    &   79 $\pm$ 3.1    \\ \hline

\end{tabular}
\end{adjustbox}
  \caption{Percentage of galaxies with stellar velocity rotation-to-dispersion ratio below 0.7 within each sample.}
  \label{anex33}
\end{table}

\begin{table}
\centering
\begin{adjustbox}{width=6.5cm}
\begin{tabular}{ccc}

\cline{2-3}
                        & \multicolumn{2}{c}{\makecell{Percentage of misaligned galaxies (\%)}} \\ \hline
\multicolumn{1}{c}{{Threshold}} & {ETG} & {LTG}  \\ \hline \hline

\multicolumn{1}{l}{$D/T$ = 0.45}  &  23 $\pm$ 0.6 &  6 $\pm$ 0.3  \\ \hline
\multicolumn{1}{l}{$v_{\rm{rot}}/\sigma_{0}$ = 0.7} & 23 $\pm$ 0.6      &  6 $\pm$ 0.3      \\ \hline
\multicolumn{1}{l}{$\kappa_{\rm{co}}/\sigma_{0}$ = 0.4} & 25 $\pm$ 0.6      &  5 $\pm$ 0.3\\ \hline
\end{tabular}
\end{adjustbox}
  \caption{Percentage of misaligned early-type and late-type galaxies predicted by different thresholds at $z=0.00-0.10$. Errors correspond to Poisson uncertainties.}\label{perclt}
\end{table}

\begin{table*}
  \includegraphics[width=16cm]{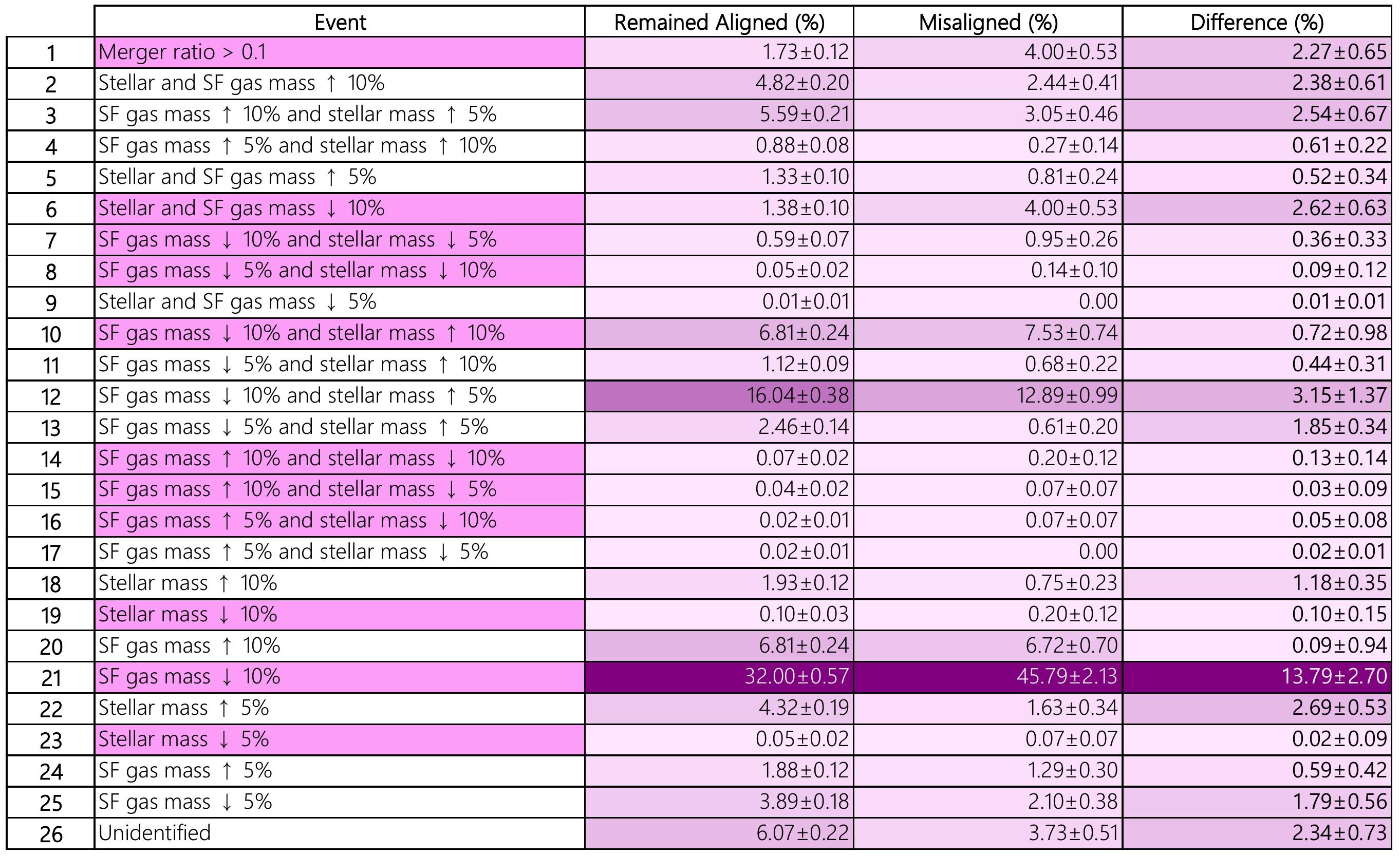}
   \caption{Percentage of galaxies `Remained Aligned' and `Misaligned' associated with the events described in Section \ref{sec:ext} at $z=0.00-0.10$ (i.e., the ratio between the number of galaxies of a population that experienced an event in the previous snapshot and the total number of galaxies in that population). The last column shows the absolute difference between the percentages of each population. The numerical values are coloured darker the higher they are, and the events coloured pink correspond to those in which the value for `Misaligned' is higher. Errors correspond to Poisson uncertainties. }
\label{tablapie}
\end{table*}


\bsp	
\label{lastpage}
\end{document}